\documentclass[8.5pt,twoside,twocolumn]{article}
\usepackage[super,sort&compress,comma]{natbib} 
\usepackage{mhchem}
\usepackage{times}
\usepackage{sectsty}
\usepackage{balance} 
\usepackage{graphicx} 
\usepackage{lastpage}
\usepackage{afterpage}
\usepackage[format=plain,justification=raggedright,singlelinecheck=false,font=small,labelfont=bf,labelsep=space]{caption} 
\usepackage{fancyhdr}
\begin{document}

\thispagestyle{plain}
\fancypagestyle{plain}{
\renewcommand{\headrulewidth}{1pt}}
\renewcommand{\thefootnote}{\fnsymbol{footnote}}
\renewcommand\footnoterule{\vspace*{1pt}%
\hrule width 3.4in height 0.4pt \vspace*{5pt}} 
\setcounter{secnumdepth}{5}

\renewcommand{\floatpagefraction}{0.35}

\renewcommand{\bottomfraction}{0.85}
\renewcommand{\textfraction}{0.05}
\renewcommand{\floatpagefraction}{0.9}

\makeatletter 
\def\subsubsection{\@startsection{subsubsection}{3}{10pt}{-1.25ex plus -1ex minus -.1ex}{0ex plus 0ex}{\normalsize\bf}} 
\def\paragraph{\@startsection{paragraph}{4}{10pt}{-1.25ex plus -1ex minus -.1ex}{0ex plus 0ex}{\normalsize\textit}} 
\renewcommand\@biblabel[1]{#1}            
\renewcommand\@makefntext[1]%
{\noindent\makebox[0pt][r]{\@thefnmark\,}#1}
\makeatother 
\renewcommand{\figurename}{\small{Fig.}~}
\sectionfont{\large}
\subsectionfont{\normalsize} 

\fancyfoot{}
\fancyfoot[RO]{\footnotesize{\sffamily{1--\pageref{LastPage} ~\textbar  \hspace{2pt}\thepage}}}
\fancyfoot[LE]{\footnotesize{\sffamily{\thepage~\textbar\hspace{3.45cm} 1--\pageref{LastPage}}}}
\fancyhead{}
\renewcommand{\headrulewidth}{1pt} 
\renewcommand{\footrulewidth}{1pt}
\newcommand{\clc}[1]{\multicolumn{1}{c}{#1}}
\setlength{\arrayrulewidth}{1pt}
\setlength{\columnsep}{6.5mm}
\setlength\bibsep{1pt}

\twocolumn[
  \begin{@twocolumnfalse}
\noindent\LARGE{\textbf{Cold collisions of an open-shell S-state atom with a $^2\Pi$ molecule:
 \\N($^4$S) colliding with OH in a magnetic field}}
\vspace{0.6cm}

\noindent\large{\textbf{Wojciech Skomorowski,\textit{$^{a}$}  
Maykel L.~Gonz\'alez-Mart\'{\i}nez,\textit{$^{b}$}
Robert Moszynski,$^{\ast}$\textit{$^{a}$} \\ and Jeremy M. Hutson$^{\ast}$\textit{$^{b}$}}}\vspace{1.5cm}


\noindent \normalsize{
We present quantum-theoretical studies of collisions between an
open-shell S-state atom and a $^2\Pi$-state molecule in the presence of
a magnetic field. We analyze the collisional Hamiltonian and discuss
possible mechanisms for inelastic collisions in such systems. The
theory is applied to the collisions of the nitrogen atom ($^4$S) with
the OH molecule, with both collision partners initially in fully
spin-stretched (magnetically trappable) states, assuming that the
interaction takes place exclusively on the two high-spin (quintet)
potential energy surfaces. The surfaces for the quintet states are
obtained from spin-unrestricted coupled-cluster calculations with
single, double, and noniterative triple excitations. We find
substantial inelasticity, arising from strong couplings due to the
anisotropy of the interaction potential and the anisotropic spin-spin
dipolar interaction. The mechanism involving the dipolar interaction
dominates for small magnetic field strengths and ultralow collision
energies, while the mechanism involving the potential anisotropy
prevails when the field strength is larger (above 100~G) or the
collision energy is higher (above 1~mK). The numerical results suggest
that sympathetic cooling
of magnetically trapped OH by collisions with ultracold 
N atoms will not be successful at higher temperatures.
}
\vspace{0.5cm}
 \end{@twocolumnfalse}
  ]

\section{Introduction}

\footnotetext{\textit{$^{a}$~Faculty of Chemistry, University of Warsaw, Pasteura 1,
02-093 Warsaw, Poland. E-mail: rmoszyns@tiger.chem.uw.edu.pl}}
\footnotetext{\textit{$^{b}$~Department of Chemistry, Durham University,
South Road, Durham DH1 3LE, United Kingdom. E-mail:J.M.Hutson@durham.ac.uk }}



The first experimental realization of Bose-Einstein condensation in a
dilute gas in 1995 \cite{bec1995} opened up a novel and fast-growing
field of research on cold and ultracold matter. At temperatures below
about $10^{-6}$~K, novel properties emerge in which the quantum nature
of atoms and molecules is crucial. Although the original experiments
involved quantum-degenerate states in atomic systems, it was soon
realised that molecules, especially those with a permanent dipole
moment, offer an additional range of applications in physics and
chemistry. These include development of new frequency standards, tests
of fundamental physical concepts such as parity and time-reversal
violation \cite{fineevolution,edm}, spectroscopic measurements of
unprecedented accuracy \cite{ybphoto,Enomoto}, quantum information
processing,\cite{Rabl,Demille}, and control of chemical reactions with
state-selected reagents and products
\cite{Kremspccp,Hutsonscience,Ospelkaus:2010}.

In contrast to atoms, which nowadays can be cooled relatively easily by
laser Doppler cooling and evaporative cooling \cite{Cohen}, molecules
are incomparably more challenging because of their complicated internal
structure. Two main classes of methods have been established to produce
cold molecules: direct methods, in which molecules are cooled from high
temperature by means of a buffer gas or external fields, and indirect
methods, in which cold molecules are formed from precooled atoms by
photoassociation or magnetoassociation.

Indirect methods can now produce ground-state molecules at temperatures
below 1 $\mu$K \cite{Ni:2008,Deiglmayr:2008,Danzl:ground:2010}. It has
been shown recently that, for KRb, the rates of chemical reactions
change spectacularly between different nuclear spin states and can be
dramatically affected by applied electric fields \cite{Ospelkaus:2010}.
However, indirect methods are so far restricted to alkali-metal dimers
and it will be challenging to extend them to other regions of the
periodic table \cite{Zuchowski:RbSr:2010}.

Direct cooling methods can be applied to a much larger variety of
chemically interesting molecules, including OH, NH$_3$, CO and LiH
\cite{Bethlem:1999,Bethlem:2000,Tokunaga:2009,Gilijamse:2006}. Stark
deceleration, pioneered by Meijer and coworkers \cite{Bethlem:1999},
can be applied to polar molecules with large Stark effects, while
helium buffer-gas cooling \cite{Weinstein:CaH:1998} has been
particularly successful for paramagnetic species. However, the
temperatures so far achieved with direct methods are limited to tens of
millikelvin, which is not cold enough to achieve quantum degeneracy.
The development of a second-stage cooling method for such molecules is
the biggest current challenge in the field. One of the most promising
proposals is to use {\em sympathetic cooling}, which is based on the
conceptually simple idea of bringing cold molecules into thermal
contact with a bath containing ultracold atoms. So far sympathetic
cooling has been successfully realized for ions
\cite{sympIon,complexIon} and some neutral atoms \cite{sympAt,sympK},
but not for neutral molecules.

Linear molecules in spatially degenerate electronic states ($\Pi$,
$\Delta$, etc.) are particularly attractive for Stark deceleration, as
they exhibit first-order Stark effects at moderate electric fields (in
contrast to molecules in $\Sigma$ states, which exhibit only
second-order Stark effects). After deceleration, the molecules can be
loaded into traps where they are confined by static electric or
magnetic fields. Such static traps are not the only way to confine cold
molecular species \cite{vanVeldhoven:2005}, but they are experimentally
the most accessible. In addition, atoms in open-shell S states (such as
alkali-metal atoms, H($^2$S), N($^4$S), He($^3$S), and Cr($^7$S)), can
be held in magnetic traps and may be suitable as coolants.

Trapping with a static field is possible only if the atom or molecule
is in a low-field-seeking state. However, the absolute ground state is
always high-field-seeking. Thus, in addition to the elastic collisions
that lead to thermalization of the sample, there is always the
possibility of inelastic collisions that transfer the colliding
partners to a lower state and release kinetic energy. Inelastic
collisions eject molecules from the trap and may lead to the heating of
the sample. The success of sympathetic cooling therefore depends on the
ratio of elastic to inelastic events, which should preferably be as
large as possible.

Molecular sympathetic cooling was first suggested for
Rb+NH($^3\Sigma^-$) \cite{Hutson:2004}. Subsequently, potential energy
surfaces and the appropriate collision cross sections have been
calculated for a variety of candidate systems, including
Mg+NH($^3\Sigma^-$) \cite{Wallis:2009}, Li+LiH($^1\Sigma^+$)
\cite{sean,skomo:LiH}, Rb+NH$_3$ \cite{pzuch:nh3} and
He+CH$_2$($^3B_1$) \cite{timur:poly}. Rb+ND$_3$ has also been explored
experimentally \cite{heather}, though the inelastic collision rate in
an electric field turned out to be too high for cooling. Studies of
cold collisions with linear molecules in a $\Pi$ state in the presence
of external fields have mostly been limited to cases when the second
colliding partner is closed-shell. In particular, Tscherbul {\em et
al.}\ \cite{timur} have investigated OH+He collisions and have shown
how the inelastic cross sections can be reduced by combining electric
and magnetic fields to eliminate certain inelastic channels. Collisions
of rotationally excited OH with He in the presence of electromagnetic
field were analyzed by Pavlovic {\em et al.} \cite{Pavlovic}, while
Bohn and coworkers \cite{Avdeenkov} studied cold collisions between two
OH molecules with long-range dipole-dipole interactions and concluded
that the evaporative cooling of OH would be challenging. Lara {\em et
al.}\ \cite{Lara} carried out theoretical studies of cold collisions of
OH with Rb, taking account of multiple potential energy surfaces and
including the hyperfine structure of OH. However, they did not include
external field effects.

There is thus a need for rigorous quantum studies of collisions between
a $\Pi$-state molecule and an open-shell S-state atom in the presence
of external fields. In this paper, we extend the theory presented in
Refs.\ \cite{Alexander:5974,Lara} to handle this case. This theory will
be applicable to a broad set of experimentally important systems,
including interactions of molecules such as OH, NO, ClO, and CH with
alkali-metal and other magnetically trappable atoms. As an example, we
present numerical results for collisions between OH($^2\Pi$) and
N($^4$S) in a magnetic field, with both colliding species initially in
their fully spin-stretched low-field-seeking states. OH was one of the
first molecules to be decelerated and trapped
\cite{Gilijamse:2006,Sawyer}, and many pioneering experiments with it
have been reported. Gilijamse {\em et al.}\ \cite{Gilijamse:2006}
carried out a crossed-beam experiment, colliding velocity-controlled OH
molecules with Xe atoms; they were able to resolve state-to-state
inelastic cross sections as a function of the collision energy. Similar
experiments with improved sensitivity have recently been performed for
OH colliding with Ar, He, and D$_2$
\cite{Kirste,Scharfenberg:2010,Scharfenberg:2011}. An experiment to
collide two velocity-controlled beams, of OH and NO, is in preparation
\cite{Meerakker}. Sawyer {\em et al.}\ \cite{Sawyer} have measured
energy-dependent cross sections for collisions between magnetically
trapped OH and slow D$_2$ molecules.

Tscherbul {\em et al.}\ \cite{timur:n} have recently suggested that
spin-polarized nitrogen atoms are a promising coolant for sympathetic
cooling experiments. N atoms at $T>$1~mK are stable against collisional
relaxation between different Zeeman levels for a wide range of magnetic
field strengths. Moreover, the low polarizability of the N atom leads
to potential energy surfaces with an anisotropy much smaller than is
usually encountered for interactions with alkali-metal atoms.
Theoretical and experimental studies for collisions of magnetically
trapped N($^4$S) and NH($^3\Sigma^-$) have been reported
\cite{pzuch:n,Hummon}, showing that the trap loss in this system is
fairly small and is caused mostly by the anisotropic magnetic
dipole-dipole interaction between the atomic and molecular spins.

This paper is organized as follows. In Sec.\ II we describe
calculations of the high-spin (quintet) potential energy surfaces
resulting from interaction of the N($^4$S) atom with the OH($^2\Pi$)
molecule. In Sec.\ III we describe the effective Hamiltonian used in
the dynamical calculations and give the expressions for the matrix
elements of the Hamiltonian. In Sec.\ IV we discuss the results of the
scattering calculations and their implications for sympathetic cooling
of OH by N atoms. Finally, Sec.\ V summarizes and concludes the paper.

\begin{figure*}[t!]
\centering
\begin{tabular}{cc}
\hspace*{-0.80cm}{\includegraphics[scale=0.47,angle=0]{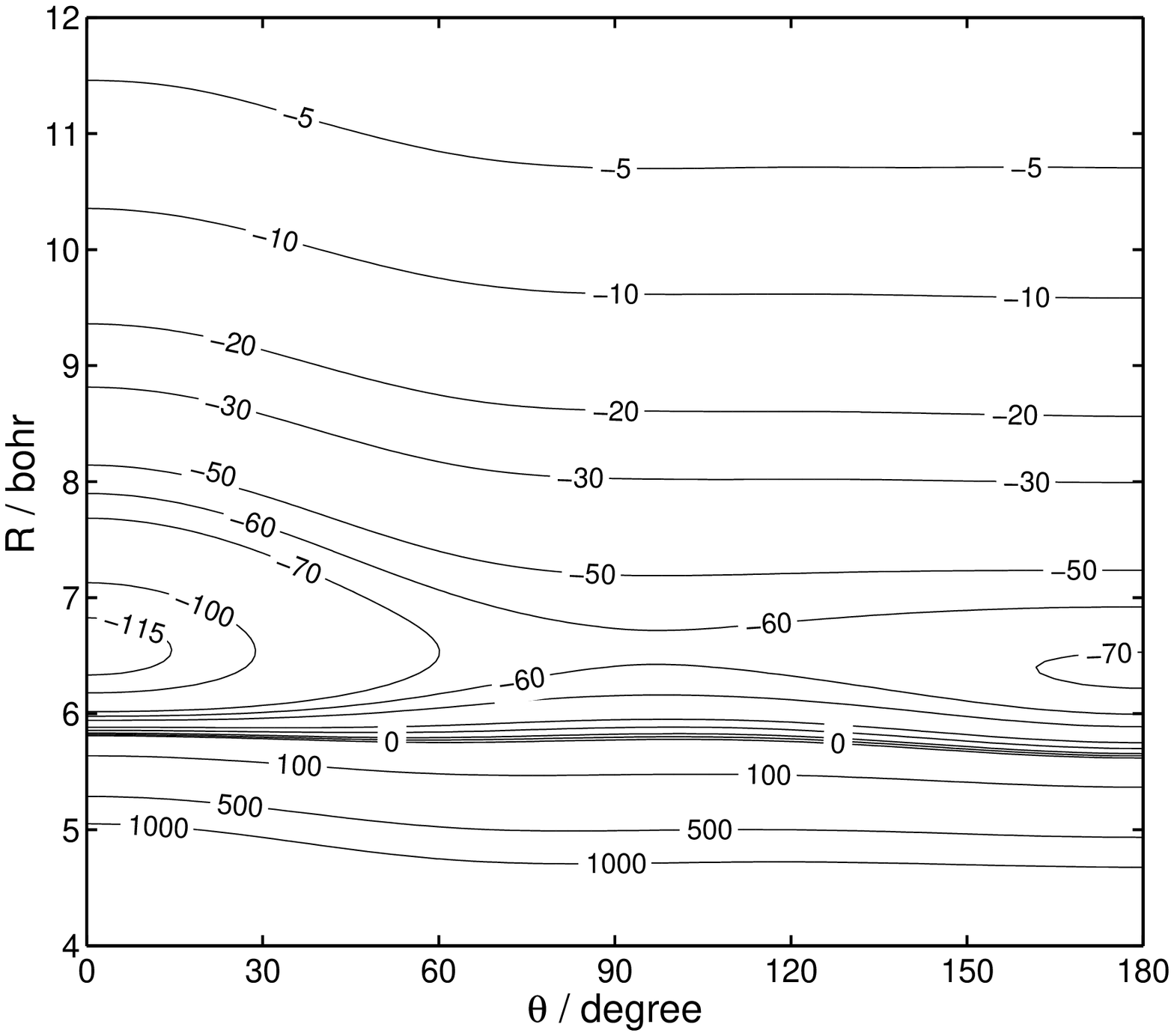}}&\hspace*{-0.80cm}
{\includegraphics[scale=0.47,angle=0]{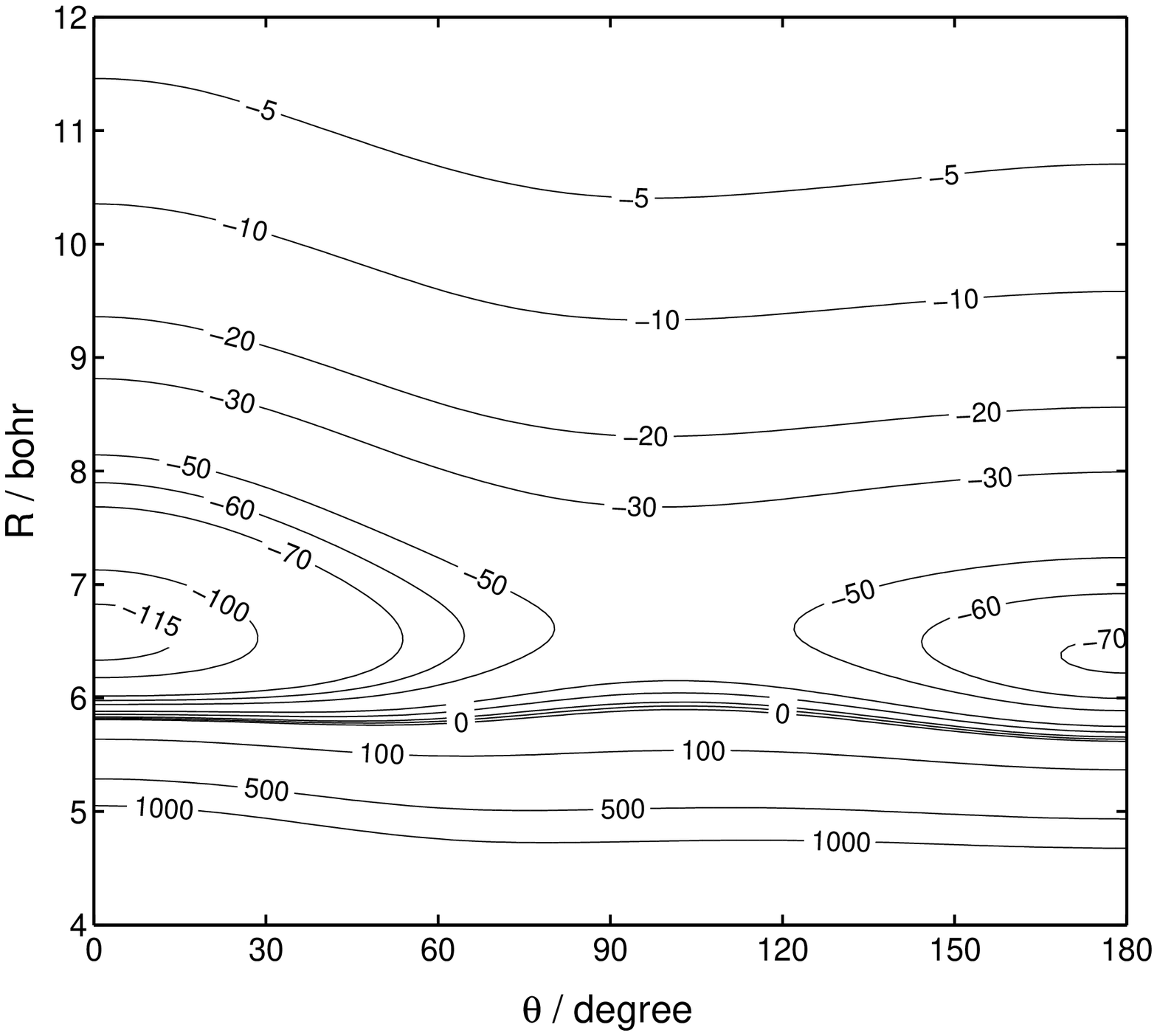}}\\
\end{tabular}
\caption{Contour plots of the quintet interaction potentials for N+OH:
$^5A^{\prime}$  (left-hand panel) and $^5A^{\prime\prime}$ (right-hand panel).
Energies are in cm$^{-1}$.}
\label{fig1}
\end{figure*}

\section{Potential energy surfaces}
The interaction between the N($^4$S) atom and the OH($^2\Pi$) molecule
occurs on four adiabatic surfaces: $^3A^\prime$, $^3A^{\prime\prime}$,
$^5A^\prime$, and $^5A^{\prime\prime}$. The triplet surfaces have been
studied extensively to investigate the reaction N+OH$\to$NO+H that can
take place on the $^3A^{\prime\prime}$ surface
\cite{Pauzat199371,sengupta:3906,guadagnini:774,guadagnini:784,
jorfi:094302,Edvardsson2006261}. This reaction is the major source of
the NO radical in the interstellar medium and is one of the key
elementary processes in nitrogen chemistry. Formation of NO is
barrierless, via a stable intermediate complex NOH, and is highly
exothermic with 1.83 eV energy release. The other possible reaction
channel N+OH$\to$NH+O is energetically forbidden for low-energy
collisions. If we neglect minor spin-orbit coupling effects between the
triplet and quintet states, the quintet surfaces are non-reactive. To
our knowledge, the quintet surfaces of N+OH have not been reported in
the literature thus far.

\begin{table}[b!]
\small
\caption{Characteristic points of the interaction potentials for the
quintet states of N($^4$S) + OH($^2\Pi$). } \label{tab1} 
  \begin{tabular*}{0.48\textwidth}{@{\extracolsep{\fill}}lcrrc}
    \hline
     & $R$ / $a_0$ & $\theta$ / degrees & $V$ / cm$^{-1}$ & Surface \\
    \hline
  Global minimum & 6.55   &  0.0$^\circ$    & --120.9     & $^5A^{\prime}$, $^5A^{\prime\prime}$ \\
  Local minimum & 6.36   & 180.0$^\circ$    & --71.5     & $^5A^{\prime}$, $^5A^{\prime\prime}$ \\
  Saddle point  & 6.56   & 97.2$^\circ$    & --61.0     & $^5A^{\prime}$          \\
  Saddle point  & 6.66   & 100.1$^\circ$    & --45.8     & $^5A^{\prime\prime}$       \\
    \hline
  \end{tabular*}
\end{table}
We have carried out calculations of the quintet surfaces using the
unrestricted version of the coupled-cluster method with single, double,
and noniterative triple excitations [UCCSD(T)]. The unrestricted
version was chosen to circumvent the problem of the lack of
size-consistency for the interaction between two open-shell systems in
spin-restricted coupled-cluster calculations \cite{liesbeth}. The
highly accurate aug-cc-pV5Z basis set of Dunning \cite{jr.:1007} was
employed for all atoms and the counterpoise procedure \cite{Boys:70}
was used to correct the computed interaction energies for basis-set
superposition error. The {\sc molpro} suite of codes \cite{molpro08}
was used in the electronic structure calculations.

Both the $^5A^\prime$ and $^5A^{\prime\prime}$ potential energy
surfaces were computed on a grid of points in Jacobi coordinates
($R,\theta$), where $R$ is the intermolecular distance measured from
the centre of mass of $^{16}$OH to the $^{14}$N atom and $\theta$ is
the angle between the vector pointing from O to H in the OH molecule
and the vector pointing from the centre of mass of OH to the N atom.
The angle $\theta=0^\circ$ thus corresponds to the linear O--H---N
arrangement. The distance $R$ was varied from 4.0 to 12.0 $a_0$ with an
interval of 0.5 $a_0$ and from 12.0 to 20.0 $a_0$ with an interval of
1.0 $a_0$. The angular grid points was chosen as the set of points for
11-point Gauss-Lobatto quadrature, which include points at $\theta=0$
and $180^\circ$. The OH bond length was kept fixed at the monomer
equilibrium value of 1.834 $a_0$.

Contour plots of the $^5A^\prime$ and $^5A^{\prime\prime}$ potential
energy surfaces are shown in Fig.\ 1. The shapes of the two quintet
potentials are quite similar. The global minima appear for the linear
geometry O--H---N and have a depth of 120.9 cm$^{-1}$. There are also
local minima 71.5 cm$^{-1}$ deep, which occur at the linear N---O--H
configuration. Note that for linear geometries the $^5A^\prime$ and
$^5A^{\prime\prime}$ states are degenerate, so these minima are common
to the two surfaces. The set of stationary points of the potentials is
completed by saddle points between the two minima, which are located at
slightly different positions for the $^5A^\prime$ and
$^5A^{\prime\prime}$ states. Table 1 gives the positions of the
stationary points on the surfaces and the corresponding interaction
energies. The shapes of the quintet potential energy surfaces for N+OH
closely resemble the high-spin (quartet) surface for N+NH reported by
\.Zuchowski and Hutson \cite{pzuch:n}, although the global minimum for
N+OH is about 30 cm$^{-1}$ deeper than for N+NH.

To perform quantum scattering calculations, it is necessary to expand
the $^5A^\prime$ and $^5A^{\prime\prime}$ surfaces in terms of
angular functions. We adopt the convention of Alexander \cite{Alexander:5974} and use
spherical harmonics in the Racah normalization $C_{k,q}(\theta,\phi)$ 
(with angle $\phi=0$) for angular representation of the potential.
For the
interaction of an S-state atom with a $\Pi$-state molecule, there are
nonvanishing terms with $q=0$ and $q=2$. The sum of the $^5A^\prime$
and $^5A^{\prime\prime}$ potentials is expanded in terms of functions 
with $k=0$,
\begin{equation}
\frac{1}{2}\left[V_{A^\prime}(R,\theta)+V_{A^{\prime\prime}}(R,\theta)\right]
= \sum_{k=0}^{\infty} C_{k,0}(\theta,0) V_{k0}(R),
\label{psum}
\end{equation}
while the difference between the $^5A^\prime$ and $^5A^{\prime\prime}$
potentials is expanded in terms of functions with $k=2$,
\begin{equation}
\frac{1}{2}\left[V_{A^\prime}(R,\theta)-V_{A^{\prime\prime}}(R,\theta)\right]
= \sum_{k=2}^{\infty} C_{k,2}(\theta,0) V_{k2}(R).
\label{pdiff}
\end{equation}
Note that the definition of
the difference potential, either $V_{A^\prime}-V_{A^{\prime\prime}}$ or
$V_{A^{\prime\prime}}-V_{A^{\prime}}$, depends in general on the
symmetry of the electronic wave functions of the interacting subsystems
\cite{nielson:2055,wskomo}. The radial functions $V_{kq}(R)$ are
obtained by projecting the sum or difference onto the appropriate
angular function, using Gauss-Lobatto quadrature to perform the
numerical integration. Prior to this projection, the interpolation to
obtain $V_{A^\prime}(R,\theta)$ and $V_{A^{\prime\prime}}(R,\theta)$ at
an arbitrary value of $R$ is done for each value of $\theta$ using the
reproducing kernel Hilbert space (RKHS) procedure \cite{ho:2584}. For
the quintet states of N($^4$S)+OH($^2\Pi$), the dominant anisotropic
term in the expansion (\ref{psum}) is $V_{20}(R)$ with a well depth of
approximately 28 cm$^{-1}$, while the dominant term in the expansion
(\ref{pdiff}) comes from $V_{22}(R)$.

\begin{table}[b!]
\small
\caption{Long-range coefficients (in atomic units) for
N($^4$S)+OH($^2\Pi$). \label{tab2}} 
  \begin{tabular*}{0.48\textwidth}{@{\extracolsep{\fill}}lrrrrrr}
    \hline
$ k ~ \rightarrow$    & \clc{0} & \clc{1} & \clc{2} & \clc{3} & \clc{4} \\
\hline
$C_6^{k0}$        &  27.84 &      & 4.92  &      &      \\
$C_6^{k2}$        &     &      & 1.23  &      &      \\
$C_7^{k0}$        &     &  51.60  &     & 24.61  &      \\
$C_7^{k2}$        &     &      &     & --6.38  &      \\
$C_8^{k0}$        & 583.34 &      & 312.00 &      & 48.29   \\
$C_8^{k2}$        &     &      & 159.09 &      & 31.42   \\
    \hline
  \end{tabular*}
\end{table}
To improve the description of the potential at large $R$, we use an
analytic representation in this region. Each radial component
$V_{kq}(R)$ is expanded at long range in terms of Van der Waals
coefficients,
\begin{equation}
V_{kq}(R)=-\sum_{n=6} C_n^{kq}/R^n.
\end{equation}
The expressions for the $C_n^{kq}$ coefficients have been given by
Skomorowski and Moszynski \cite{wskomo}, though with a different normalisation 
for $q>0$ from the one used here. 
We calculated the Van der
Waals constants up to and including $n$=8, using the method described
in Ref.\ \cite{wskomo}. The results are listed in Table 2. 
For a weakly
polarizable system such as N+OH, the neglect of higher-order
coefficients with $n>8$ is fully justified. We used the switching
function of Janssen {\em et al.} \cite{liesbeth}, with parameters
$a=15$ $a_0$ and $b=25$ $a_0$, to join the asymptotic form based on the
long-range coefficients and the RKHS interpolation of the {\it ab
initio} points. \label{sec1}

\section{Collision Hamiltonian}

\subsection{Effective Hamiltonian}
We consider the case of an atom A($^{2s_1+1}$S), interacting with a
diatomic molecule BC($^{2}\Pi$), in the presence of an external
magnetic field $B$. The direction of the field defines the laboratory
(space-fixed) $Z$-axis. The system A--BC is described in Jacobi
coordinates, with the $\boldsymbol{r}$ vector connecting the heavier
and lighter of the atoms B and C, and $\boldsymbol{R}$ connecting the
centre of mass of BC and the atom A. By convention, lower-case and
capital letters are used to represent the quantum numbers of the
monomers and of the complex as a whole, respectively. The subscripts 1
and 2 refer to the monomers A and BC, respectively. For simplicity, the
diatom will be treated as a rigid rotor in vibrational state $v$,
although generalization to include its vibrations is straightforward.

The Hamiltonian describing the nuclear motions of A+BC in the presence
of magnetic field $B$ can be written
\begin{equation}
 \hat{H} = - \frac{\hbar^2}{2\mu} R^{-1} \frac{d^2}{dR^2} R
           + \frac{\hat{L}^2}{2\mu R^2}
           + \hat{H}_\mathrm{mon}
           + \hat{H}_{12},
 \label{eq:Heff}
\end{equation}
where $\hat{L}$ is the space-fixed angular momentum operator describing
the end-over-end rotation of A and BC about one another and $\mu$ is
the reduced mass of the complex. $\hat{H}_\mathrm{mon}$ contains all
terms describing the \emph{isolated} monomers, i.e.
$\hat{H}_\mathrm{mon} = \hat{H}_\mathrm{1}
 + \hat{H}_\mathrm{2}$. $\hat{H}_{12}$
describes the interaction between the monomers:
\begin{equation}
 \hat{H}_{12} = \hat{H}_\mathrm{s_1s_2} +
\hat{V}(R,\theta).
\end{equation}
Here, $\hat{H}_\mathrm{s_1s_2}$ accounts for the direct dipolar
interaction between the magnetic moments due to the unpaired electrons
of the monomers, and $\hat{V}$ is the intermolecular interaction
potential.

If $s_1 \ne 0$ and hyperfine terms are neglected, the Hamiltonian for
an isolated atom in the state $^{2s_1+1}{\rm S}$ is fully determined by
the Zeeman interaction between the electron spin and the external
magnetic field,
\begin{equation}
 \hat{H}_\mathrm{1} =
 g_S \mu_\mathrm{B}\;
 \hat{s}_1 \cdot \hat{ B},
 \label{eq:Hmon1}
\end{equation}
where $g_S$ is the electron $g$-factor, $\mu_\mathrm{B}$ is the
electron Bohr magneton, and $\hat{s}_1$ is the spin operator.

The analogous Hamiltonian for a $^2\Pi$ molecule can be written
\cite{jmbrown:03}
\begin{equation}
\hat{H}_\mathrm{2} =
\hat{H}_\mathrm{rso} + \hat{H}_\mathrm{Z,2} + \hat{H}_\lambda,
\label{hmon2}
\end{equation}
where the rotational and spin-orbit contributions within the $\Pi$
state are collapsed into the first term,
\begin{equation}
\hat{H}_\mathrm{rso} \equiv B_v \, \hat{n}^2
+ A_v \;\hat{l} \cdot \hat{s_2}.
\end{equation}
$B_v$ and $A_v$ are the molecular rotational and spin-orbit constants,
respectively, and $\hat{n}$ is the operator of the mechanical rotation
of BC, which can be expressed as $\hat{\jmath} - \hat{l} - \hat{s}_2$,
where $\hat{\jmath}$, $\hat{l}$ and $\hat{s}_2$ are the operators for
the rotational, electronic orbital and spin angular momenta,
respectively. $\hat{H}_\mathrm{rso}$ can be rewritten
\begin{eqnarray}
 \hat{H}_\mathrm{rso}
     &=& (A_v+2B_v) \,\hat{l}_z \hat{s}_{2z}
       \cr && + B_v \Bigr[
       \hat{\jmath}^2
       + \hat{l}^2
       + \hat{s}_2^2
   - 2 \hat{\jmath} \cdot \hat{s}_2
         + \hat{l}_z^2
         - 2 \hat{\jmath}_z \, \hat{l}_z \Bigr].
 \label{eq:Hrso}
\end{eqnarray}
The terms $\hat{l}^2$, $\hat{s}_2^2$ and $\hat{l}_z^2$ simply shift all
the levels by a constant amount and are omitted below. The term
$\hat{H}_\lambda$, responsible for the $\Lambda$-doubling of the
rotational levels of BC, is represented by the effective Hamiltonian
\begin{equation}
 \hat{H}_\lambda = \sum_{q=\pm 1} \mathrm{e}^{-2\mathrm{i}q\phi_r}
     \left[ 
        -q_v \mathrm{T}^2_{2q}(\hat{\jmath},\hat{\jmath})+ (p_v+2q_v) \mathrm{T}^2_{2q}(\hat{\jmath},\hat{s}_2) \right],
 \label{eq:Hlambda}
\end{equation}
where $\phi_r$ is the azimuthal angle associated with the electron
orbital angular momentum about the molecular axis defined by
$\boldsymbol{r}$, while $p_v$ and $q_v$ are empirical parameters. In
Eq.\ (\ref{eq:Hlambda}), the second-rank tensor $\textrm{T}^2_q$ that
couples two vectors $\mathbf{k}_1$ and $\mathbf{k}_2$ is defined as
\begin{equation}
\textrm{T}^2_q(\mathbf{k}_1,\mathbf{k}_2)=\sum_{q_1,q_2}\langle1,q_1;1,q_2|2,q\rangle
\textrm{T}^1_{q_1}(\mathbf{k}_1)\,\textrm{T}^1_{q_2}(\mathbf{k}_2),
\end{equation}
where $\langle1,q_1;1,q_2|2,q\rangle$ is a Clebsch-Gordan coefficient
and the first-rank tensor components are $\textrm{T}^1_{0}(\mathbf{k})
= \textrm{k}_z$ and $\textrm{T}^1_{\pm 1}(\mathbf{k})=\mp
(\textrm{k}_x\pm i\,\textrm{k}_y)/\sqrt{2}$. If only the electron spin
and orbital contributions are taken into account, the Zeeman term is
\begin{equation}
 \hat{H}_\mathrm{Z,2} = g_S \mu_\mathrm{B} \;
         \hat{s}_2 \cdot \hat{B}
                 + g'_L \mu_\mathrm{B}\;
          \hat{l} \cdot \hat{B},
 \label{eq:HZ2}
\end{equation}
where $g'_L$ is the orbital $g$-factor. For diatomic molecules of
multiplicity higher than 2 (for example $^3\Pi$), an additional term
describing the intramolecular spin-spin interaction must be included in
the monomer Hamiltonian (\ref{hmon2}).

The spin-spin dipolar interaction can conveniently be written
\cite{jmbrown:03}:
\begin{equation}
 \hat{H}_\mathrm{s_1s_2} = -g^2_S \mu^2_B (\mu_0/4\pi) \sqrt{6}
 \sum_{q}(-1)^q\,\mathrm{T}^2_{q}(\hat{s}_1,\hat{s}_2) \, \mathrm{T}^2_{-q}(\boldsymbol{C}),
 \label{eq:Hs1s2}
\end{equation}
with $\mathrm{T}^2_q(\boldsymbol{C}) = C_{2,q}(\theta,\phi) R^{-3}$,
where $C_{2,q}(\theta,\phi)$ is a spherical harmonic function in the Racah
normalization and ($R,\theta,\phi$) is the set of relative spherical
coordinates of the `composite' atomic and diatomic electronic spins in
the space-fixed frame. $\mu_0$ is the magnetic permeability of the
vacuum.

\subsection{Basis sets and matrix elements}
\label{sec2:basis+aHa'} The state of the BC molecule can conveniently
be described using Hund's case $(a)$ basis functions $\left|\lambda;
s_2 \sigma_2; j \omega m_j\right> $, where $s_2$ is the electron spin
with projection $\sigma_2$ onto the molecular axis (body-fixed $z$
axis), $ \lambda$ is the (signed) projection of the electronic orbital
angular momentum onto the molecular axis, and $j$ is the angular
momentum of BC with projections $\omega$ onto the molecular axis and
$m_j$ onto the space-fixed $Z$-axis. For the body-fixed projections we
have $\omega=\lambda+\sigma$. The state of the atom is characterized by
the electronic spin function $\left|s_1 m_{s_1}\right>$. The basis set
used here for the A--BC collision system is constructed as $\left|s_1
m_{s_1}\right>\left|\lambda; s_2 \sigma_2; j \omega m_j\right> \left|L
M_L\right>$, where $ \left|L M_L\right>$ are functions describing the
relative motion of A and BC in the space-fixed reference frame.

In the presence of a magnetic field, the conserved quantities are the
projection $M_{\rm tot}$ of the total angular momentum, $M_{\rm
tot}=m_{s_1}+m_{j}+M_L$, and the total parity $\mathcal{P}$. An electric
field would mix states of different total parity. In the absence of an
electric field it is most efficient to use a parity-adapted basis set,
$\left|s_1 m_{s_1}\right> \left|s_2; j \bar{\omega} m_j \epsilon\right>
\left|L M_L\right>$, with
\begin{equation}
\begin{split}
 \left|s_2; j \bar{\omega} m_j \epsilon\right> \equiv& 
              \frac{1}{\sqrt{2}} \Bigr[
     \left|1; s_2 \sigma_2; j \bar{\omega} m_j\right>
  \\&+ \epsilon (-1)^{j-s_2}
    \left|-1; s_2\ {-\sigma_2}; j\ {-\bar{\omega}} m_j\right>
            \Bigr],
 \label{eq:s2jwmje}
\end{split}
\end{equation}
where $\bar{\omega} \equiv |\omega|$, $\sigma_2=\bar{\omega}-1$ and
$\epsilon = \pm 1$. In this basis set, the parity of BC is $p_2 =
\epsilon (-1)^{j-s_2}$, and that of the triatomic system is
$\mathcal{P} = p_1 p_2 (-1)^{L}$. The matrix elements of $\hat{L}^2$
and $\hat{H}_\mathrm{1}$ are diagonal, and given by $\hbar^2 L(L+1)$
and $g_S \mu_\mathrm{B} m_{s_1}\, B$, respectively.

We next give the matrix elements of all terms in the Hamiltonian of
Eq.\ (\ref{eq:Heff}), although only those involving the atomic spin are
new in the present work. The terms that do not involve atomic spin are
the same as for collisions with a closed-shell atom and were previously
given by Tscherbul {\em et al.}\ \cite{timur}. However, the published
version of Ref.\ \cite{timur} contains a number of typographical
errors, so we report the correct expressions here.

The matrix elements of the molecular rotation/spin-orbit operator are
 \begin{eqnarray}
 &&\!\! \left<s_2; j \bar{\omega} m_j \epsilon\right|
  \hat{H}_\mathrm{rso}
  \left|s_2; j \bar{\omega}' m_j \epsilon\right>
   \cr && \;\;\;\;=\delta_{\bar{\omega} \bar{\omega}'}
   \Bigl\{ (A_v+2B_v) (\bar{\omega}-1)+ B_v \left[j(j+1) - 2\bar{\omega}^2\right]
     \Bigr\}
  \cr &&
   \;\;\;\;\;\;\;\; -B_v \Bigr[ \delta_{\bar{\omega} \bar{\omega}'-1}
    \alpha_-(j,\bar{\omega}') \alpha_-(s_2,\bar{\omega}'-1)
      \cr && \;\;\;\;\;\;\;\;\;\;\;\;\;\;\;\;\; +\delta_{\bar{\omega} \bar{\omega}'+1}
    \alpha_+(j,\bar{\omega}') \alpha_+(s_2,\bar{\omega}'-1) \Bigr],
 \label{eq:aHrso-Hlambdaa'}
 \end{eqnarray}
where we use $\alpha_\pm(j,m) \equiv \sqrt{j(j+1) - m(m\pm1)}$ both to
simplify the equations and to ease comparison with Ref.\ \cite{timur}.
The off-diagonal terms on the right-hand side connect different
spin-orbit manifolds related by $\bar{\omega}' = \bar{\omega} \pm 1$.

The $\Lambda$-doubling matrix elements are
 \begin{equation}
\begin{split}
 &\hspace{-3mm}\left<s_2; j \bar{\omega} m_j \epsilon\right| \hat{H}_\lambda
  \left|s_2; j \bar{\omega}' m_j \epsilon\right>
  = \frac{1}{2} \epsilon (-1)^{j-s_2} \alpha_-(j,\bar{\omega}')
  \\ \times &\bigr[ \delta_{\bar{\omega} 2-\bar{\omega}'} q_v \alpha_-(j,\bar{\omega}'-1)
  - \delta_{\bar{\omega} 1-\bar{\omega}'} (p_v+2q_v) \alpha_+(s_2,\bar{\omega}'-1)
  \bigr]
\end{split}
 \label{eq:aHlambdaa'}
 \end{equation}
and also couple states with different $\bar{\omega}$. For a $^2\Pi$
molecule, the first factor inside the square brackets mixes the $1/2$
and $3/2$ states, while the second is non-zero only for $\bar{\omega} =
\bar{\omega}' = 1/2$.

 The matrix elements of the Zeeman interaction for the molecule BC are
{\small
 \begin{equation}
  \begin{split}
  &\hspace{-3mm}\left<s_2; j \bar{\omega} m_j \epsilon\right| \hat{H}_\mathrm{Z,2}
  \left|s_2; j' \bar{\omega}' m_j \epsilon\right>
    \\& \hspace{-3mm}=\mu_\mathrm{B} B (-1)^{m_j - \bar{\omega}'} \left[(2j+1)(2j'+1)\right]^{1/2}
  \left(\begin{array}{ccc} j  & 1 & j' \\
              m_j & 0 & -m_j \end{array}\right) 
   \\&
   \times \Bigg[ g_S \frac{\alpha_+(s_2,\bar{\omega}'-1)}{\sqrt{2}}
  \left(\begin{array}{ccc} j      & 1 & j'       \\
              \bar{\omega} & -1 & -\bar{\omega}' \end{array}\right)
   - g_S \frac{\alpha_-(s_2,\bar{\omega}'-1)}{\sqrt{2}}
  \\& 
   \times\left(\begin{array}{ccc} j      & 1 & j' \\      
              \bar{\omega} & 1 & -\bar{\omega}' \end{array}\right) 
  + \left[ g_S (\bar{\omega}-1) + g'_L \right]
  \left(\begin{array}{ccc} j      & 1 & j'       \\
              \bar{\omega} & 0 & -\bar{\omega}' \end{array}\right)
  \Bigg],
  \end{split}
 \label{eq:aHZ2a'}
 \end{equation}}
and mix both different rotational and different spin-orbit states.

To determine the matrix elements of the spin-spin dipolar interaction
it is natural to expand the second-rank tensor
$\mathrm{T}^2(\hat{s}_1,\hat{s}_2)$ as a linear combination of the
products of the space-fixed components of first-rank tensors
$\mathrm{T}^1_{p_1}(\hat{s}_1)$ and $\mathrm{T}^1_{p_2}(\hat{s}_2)$.
The matrix elements of $\mathrm{T}^1_{p_1}(\hat{s}_1)$ can be
calculated directly in our basis set, while for
$\mathrm{T}^1_{p_2}(\hat{s}_2)$ we first need to transform from the
space- to the body-fixed frame,
\begin{equation}
\mathrm{T}^1_{p_2}(\hat{s}_2) = \sum_{q}
 \mathcal{D}^{(1)*}_{p_2 q}(\Omega) \mathrm{T}^1_{q}(\hat{s}_2),
\end{equation}
where $\mathcal{D}^J_{K M}$ is a Wigner rotation matrix and $\Omega$
represents the Euler angles for the transformation. The matrix elements
in the primitive basis set are
{\small
 \begin{equation}
 \begin{split} 
  &\hspace{-2mm} \left<L M_L
  ;\lambda; s_2 \sigma_2; j \omega m_j
  ;s_1 m_{s_1}\hspace{-0.5mm}\right|\hspace{-0.5mm}
  \hat{H}_\mathrm{s_1s_2}\hspace{-1mm}
  \left|s_1 m'_{s_1}
  ;\lambda; s_2 \sigma'_2; j' \omega' m'_j
  ;L' M'_L\right>
  \\&\hspace{-2mm}  
  = -\sqrt{30} \lambda_\mathrm{s_1s_2}(R)
  (-1)^{s_1 - m_{s_1} + s_2 - \sigma_2 + m_j - \omega - M_L}
   \\&\hspace{-0mm} 
  \times \left[s_1(s_1+1)s_2(s_2+1) (2s_1+1)(2s_2+1) (2j+1)(2j'+1)
    \right]^{1/2}
  \\&\hspace{-0mm} 
  \times \left[(2L+1)(2L'+1)\right]^{1/2}\left(\begin{array}{ccc} L & 2 & L' \\
              0 & 0 & 0 \end{array}\right)
  \\&\hspace{-0mm}\times\hspace{-0mm} 
  \sum_{p_1,p_2,q} \hspace{-0mm}\left(\begin{array}{ccc} 1  & 1  & 2 \\
              p_1 & p_2 & -p \end{array}\right)
  \left(\begin{array}{ccc}   s_1 & 1  & s_1   \\
              -m_{s_1} & p_1 & m'_{s_1} \end{array}\right)
  \\&\hspace{-0mm} 
   \times \left(\begin{array}{ccc}   s_2  & 1 & s_2    \\
              -\sigma_2 & q & \sigma'_2 \end{array}\right)
  \left(\begin{array}{ccc} j  & 1  & j'  \\
              -m_j & p_2 & m'_j \end{array}\right)
  \\&\hspace{-0mm} 
  \times \left(\begin{array}{ccc} j    & 1 & j'   \\
              -\omega & q & \omega' \end{array}\right)
  \left(\begin{array}{ccc} L  & 2  & L'  \\
              -M_L & -p & M_L' \end{array}\right),
 \label{eq:aHs1s2a'primitive}
 \end{split}
 \end{equation}}
and the corresponding matrix elements in the parity-adapted basis set
are
{\small
 \begin{equation}
  \begin{split}
 &\hspace{-0mm} \left<L M_L
  ;s_2; j \bar{\omega} m_j \epsilon;s_1 m_{s_1}\right|
  \hat{H}_\mathrm{s_1s_2}
  \left|s_1 m'_{s_1};s_2; j' \bar{\omega}' m'_j \epsilon
  ;L' M'_L\right> 
  \\ & 
  =\sqrt{30} \lambda_\mathrm{s_1s_2}(R)
  (-1)^{s_1 - m_{s_1} + s_2 + m_j + 2\bar{\omega} - M_L} 
  \\&\hspace{3mm}
   \times \left[s_1(s_1+1)(2s_1+1) s_2(s_2+1)(2s_2+1) (2j+1)(2j'+1)
    \right]^{1/2}
 \\ &\hspace{3mm} 
  \times \left[(2L+1)(2L'+1)\right]^{1/2} \left(\begin{array}{ccc} L & 2 & L' \\
              0 & 0 & 0 \end{array}\right)
   \\ & \hspace{3mm}\times 
   \sum_{p_1,p_2,q} 
  \left(\begin{array}{ccc} 1  & 1  & 2 \\
              p_1 & p_2 & -p \end{array}\right)
  \left(\begin{array}{ccc}   s_1 & 1  & s_1   \\
              -m_{s_1} & p_1 & m'_{s_1} \end{array}\right)
  \\ &  \hspace{3mm} \times
  \left(\begin{array}{ccc} s_2    & 1 & s_2       \\
            -\bar{\omega}+1 & q & \bar{\omega'}-1 \end{array}\right)
  \left(\begin{array}{ccc} j  & 1  & j'  \\
              -m_j & p_2 & m'_j \end{array}\right)
  \\ &  \hspace{3mm} \times
  \left(\begin{array}{ccc} j      & 1 & j'      \\
              -\bar{\omega} & q & \bar{\omega'} \end{array}\right)
  \left(\begin{array}{ccc} L  & 2  & L'  \\
              -M_L & -p & M'_L \end{array}\right),
 \end{split}
 \label{eq:aHs1s2a'}
 \end{equation}}
where $p \equiv p_1 + p_2$, $\lambda_\mathrm{s_1s_2}(R)=E_{\rm
h}a_0^3\alpha^2/R^3$ is the $R$-dependent spin-spin dipolar coupling
constant and $\alpha\approx1/137$ is the fine-structure
constant.

Finally, the matrix elements of the interaction potential are
 \begin{equation}
 \begin{split}
 &\hspace{-0mm} \left<L M_L
  ;s_2; j \bar{\omega} m_j \epsilon\right|
  \hat{V}
  \left|s_2; j' \bar{\omega}' m'_j \epsilon';
  L' M'_L\right>
   \\ &
   =(-1)^{m_j - \bar{\omega}' - M_L}
    \left[(2j+1)(2j'+1) (2L+1)(2L'+1)\right]^{1/2} 
  \\ & \hspace{3mm} \times
  \sum_{k,m_k} \frac{1}{2}\left[1+\epsilon\epsilon'(-1)^k \right] (-1)^{m_k}
  \left(\begin{array}{ccc} j  & k  & j'  \\
              m_j & m_k & -m'_j \end{array}\right)
  \\ &  \hspace{3mm} \times
  \left(\begin{array}{ccc} L & k & L' \\
              0 & 0 & 0 \end{array}\right)
  \left(\begin{array}{ccc} L  & k  & L'  \\
              -M_L & m_k & M'_L \end{array}\right) 
  \\ &  \hspace{3mm} \times
   \Bigl[
  \left(\begin{array}{ccc} j      & k & j'       \\
              \bar{\omega} & 0 & -\bar{\omega}' \end{array}\right)
  V_{k0}(R)
   \\ &
  \hspace{25mm} 
   - (1 - \delta_{\bar{\omega} \bar{\omega}'})\epsilon' 
  \left(\begin{array}{ccc} j      & k & j'      \\
              \bar{\omega} & -2 & \bar{\omega}' \end{array}\right)
  V_{k2}(R) \Bigr],
\end{split}
 \label{eq:aVa'}
 \end{equation}
where $V_{k0}(R)$ and $V_{k2}(R)$ are the radial strength functions of
Eqs.\ (\ref{psum}) and (\ref{pdiff}). 
It is readily seen that states
belonging to the same spin-orbit manifold are coupled through the
`average' of the $A'$ and $A''$ potential surfaces, while those of
different manifolds are connected through their difference. In
addition, the factor $\frac{1}{2} \left[1+\epsilon\epsilon'(-1)^k
\right]$ guarantees that states of the \emph{same} monomer parity are
connected by terms $V_{kq}(R)$ with \emph{even} $k$, while those with
\emph{odd} $k$ couple rotational levels of \emph{opposite} parity. It
follows from this that a strong parity-conserving propensity rule for
transitions involving different spin-orbit manifolds can be expected.

\section{Dynamical calculations}

\subsection{Computational details}

Expanding the Schr\"odinger equation with the Hamiltonian of Eq.\
(\ref{eq:Heff}) in the parity-adapted basis set (\ref{eq:s2jwmje})
yields a set of coupled differential equations. We have written a
plug-in for the {\sc MOLSCAT} general-purpose quantum molecular
scattering package \cite{molscat} to implement the matrix elements
described above for collisions between an open-shell S-state atom and a
$^2\Pi$-state molecule in a magnetic field. We solved the coupled
equations numerically using the hybrid propagator of Alexander and
Manolopoulos \cite{alexander:2044}, propagating from $R_{\rm min}=4\
a_0$ to $R_{\rm mid}=25\ a_0$ using a fixed-step log-derivative
propagator with interval size 0.02 $a_0$  and from
$R_{\rm mid}$ to $R_{\rm max}=800\ a_0$ using a variable-step
log-derivative propagator based on Airy functions. {\sc MOLSCAT}
applies scattering boundary conditions at $R_{\rm max}$ to extract
scattering S-matrices, which are then used to calculate elastic and
inelastic cross sections.

Values of the OH molecular constants in the monomer Hamiltonian were
taken from Refs.\ \cite{Ticknor:OHmag:2005,Brown:OH:1978}. After
performing numerous test calculations, we decided to include basis
functions with $j\le9/2$ and $L\le8$, which gives convergence of the
cross sections to within approximately 1\%.

\subsection{Results}
\begin{figure}[t!]
{\hspace*{-2.3cm}\includegraphics[scale=0.50,angle=-90]{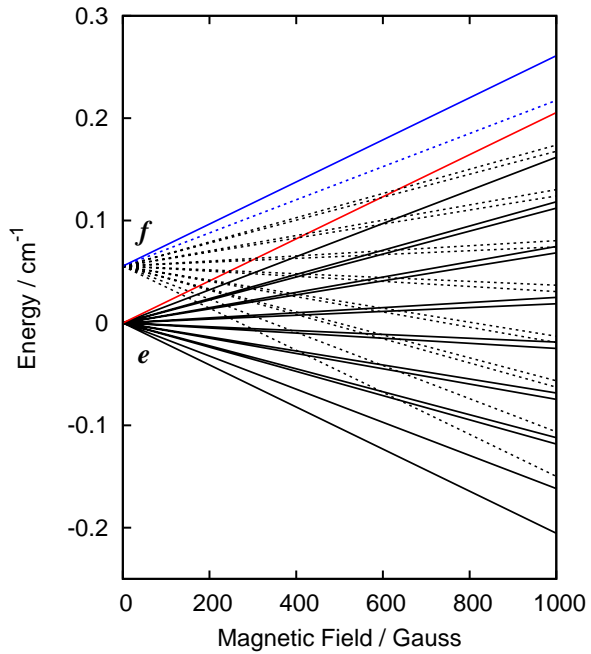}}\\
\caption{
Energy levels of noninteracting N($^4$S)+OH($X^2\Pi,j=3/2$) in a magnetic field.
The solid red and blue lines indicates the
spin-stretched low-field-seeking states $|m_{s_1}=3/2\rangle|m_j=3/2,e\rangle$
(red) and $|m_{s_1}=3/2\rangle|m_j=3/2,f\rangle$ (blue). The dotted blue
line shows state $|m_{s_1}=3/2\rangle|m_j=1/2,f\rangle$.
}
\label{fig2}
\end{figure}

The lowest rotational state of OH in its ground $X^2\Pi$ state at zero
field is a $\Lambda$ doublet with $j=3/2$. The doublet consists of two
states, referred to as $e$ and $f$, which have opposite parity and are
separated by 0.059 cm$^{-1}$, with { $|j=3/2, e\rangle$} being the
ground state. A magnetic field splits each component of the doublet
into four states differing by the projection of the angular momentum
$m_j$ on the field axis ($m_j=3/2,1/2,-1/2,-3/2$). For the N atom in
its $^4$S ground state, a magnetic field produces four Zeeman levels,
with spin projections $m_{s_1}=3/2$, 1/2, $-1/2$ and $-3/2$. The
combination of 8 Zeeman levels of OH with 4 of the N atom yields 32
asymptotic levels (thresholds), as shown in Fig.\ \ref{fig2}. In
principle, even at zero field each of the levels is further split due
to hyperfine interactions, although in the present work hyperfine
effects are neglected for simplicity.

N and OH can both be magnetically trapped in their spin-stretched
states, with $m_{s_1}=3/2$ and $m_j=3/2$, respectively. There are two
such states for OH, originating from the $e$ and $f$ components of the
$\Lambda$ doublet. We choose the initial state to be
$|m_{s_1}=3/2\rangle|m_j=3/2,e\rangle$, shown with a red line in Fig.\
\ref{fig2}. This is likely to be more favourable for sympathetic
cooling than $|m_{s_1}=3/2\rangle|m_j=3/2,f\rangle$ (shown with a solid blue
line in Fig.\ \ref{fig2}), because there are fewer inelastic channels open for
Zeeman relaxation at low collision energies. In particular, we avoid
transitions between the two fully spin-stretched states, from
$|m_{s_1}=3/2\rangle|m_j=3/2,f\rangle$ to
$|m_{s_1}=3/2\rangle|m_j=3/2,e\rangle$, at collision energies below
about 85 mK. The only centrifugal suppression in such a process, even
for an incoming $s$ wave ($L_i=0,M_{L,i}=0$) is due to a $p$-wave
barrier in the outgoing channel ($L_f=1,M_{L,f}=0$) with a height of
only 11 mK, necessitated by the change in OH monomer parity.

The interaction between collision partners that are initially in fully
spin-stretched states takes place almost entirely on the quintet
(high-spin) potential energy surfaces. A full description of exit
channels in which $m_{s_1}+m_j$ has changed requires triplet surfaces,
but including these explicitly would be computationally prohibitively
expensive. In the present work, we effectively approximate the triplet
potential surfaces with the corresponding quintet ones. This
approximation is closely analogous to that used for N+NH in ref.\
\cite{pzuch:n}.

In a low-energy inelastic collision, the quantum state of at least one
of the colliding species changes and kinetic energy is released. There
are two main mechanisms that produce inelasticity in ultracold
collisions of an open-shell S-state atom with a molecule in a $^2\Pi$
state. The first is direct coupling through the anisotropy of the
interaction potential, which drives transitions to states with the
molecular quantum number $m_{j}$ reduced by at least 1 and the atomic
spin projection $m_{s_1}$ unchanged. This mechanism is also present in
collisions between a closed-shell atom and a $^2\Pi$ molecule and has
been described by Tscherbul {\em et al.}\ \cite{timur}. The second
mechanism arises from coupling by the spin-spin dipolar interaction
$\hat{H}_\mathrm{s_1s_2}$. Here, the final Zeeman state may have
quantum numbers $m_{j}$ and $m_{s_1}$ reduced by at most one. Such
processes are also present in collisions of an open-shell S-state atom
with $^2\Sigma$ or $^3\Sigma$ molecules, or indeed between two
alkali-metal atoms. Collisions of spin-polarized S-state atoms with
$^2\Pi$ molecules thus combine two direct mechanisms for coupling
between different Zeeman levels.

The most important contribution to coupling by the interaction
potential comes from the anisotropic term $V_{20}(R)$, which induces
direct transitions from the OH state $|m_j=3/2,e\rangle$ to
$|m_j=1/2,e\rangle$ and $|m_j=-1/2,e\rangle$. This occurs even in the
$s$-wave regime ($L_i=0$). An $s$-wave collision in which $m_{s_1}+m_j$
decreases requires $M_{L,f}>1$ to conserve $M_{\rm tot}$. If the monomer
parity is unchanged, conservation of total parity then requires
$L_f\ge2$. There is thus a centrifugal barrier in the outgoing channel,
which suppresses the inelastic cross sections for low collision
energies and low fields. For N+OH, the centrifugal barriers are
relatively high due to the low reduced mass and small $C_{6}^{00}$
coefficient: the height of the $d$-wave barrier is 71 mK.
\begin{figure}[t!]
\vspace*{0cm}
{\hspace*{-0.4cm}\includegraphics[scale=0.37,angle=-90]{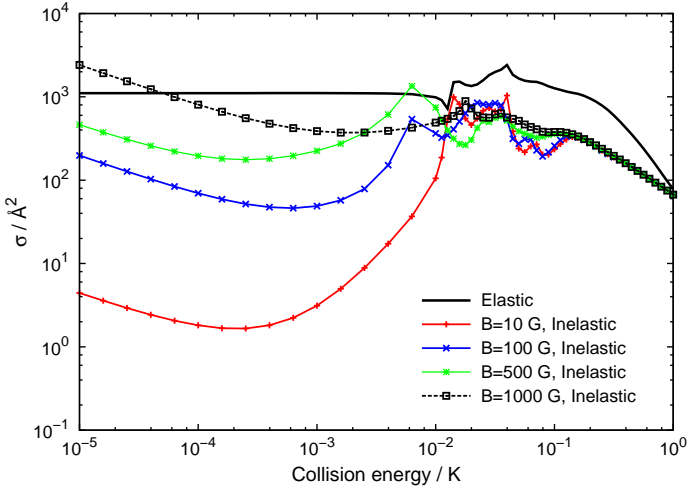}}\\
\caption{
Elastic and total inelastic cross sections for N+OH scattering at
different magnetic field strengths $B$. The elastic cross section is
almost unaffected by the field strength and is shown only for $B=10$ G.
} \label{fig3}
\end{figure}

\begin{figure}[b!]
\centering
\hspace*{-0.20cm}{\includegraphics[scale=0.32,angle=-90]{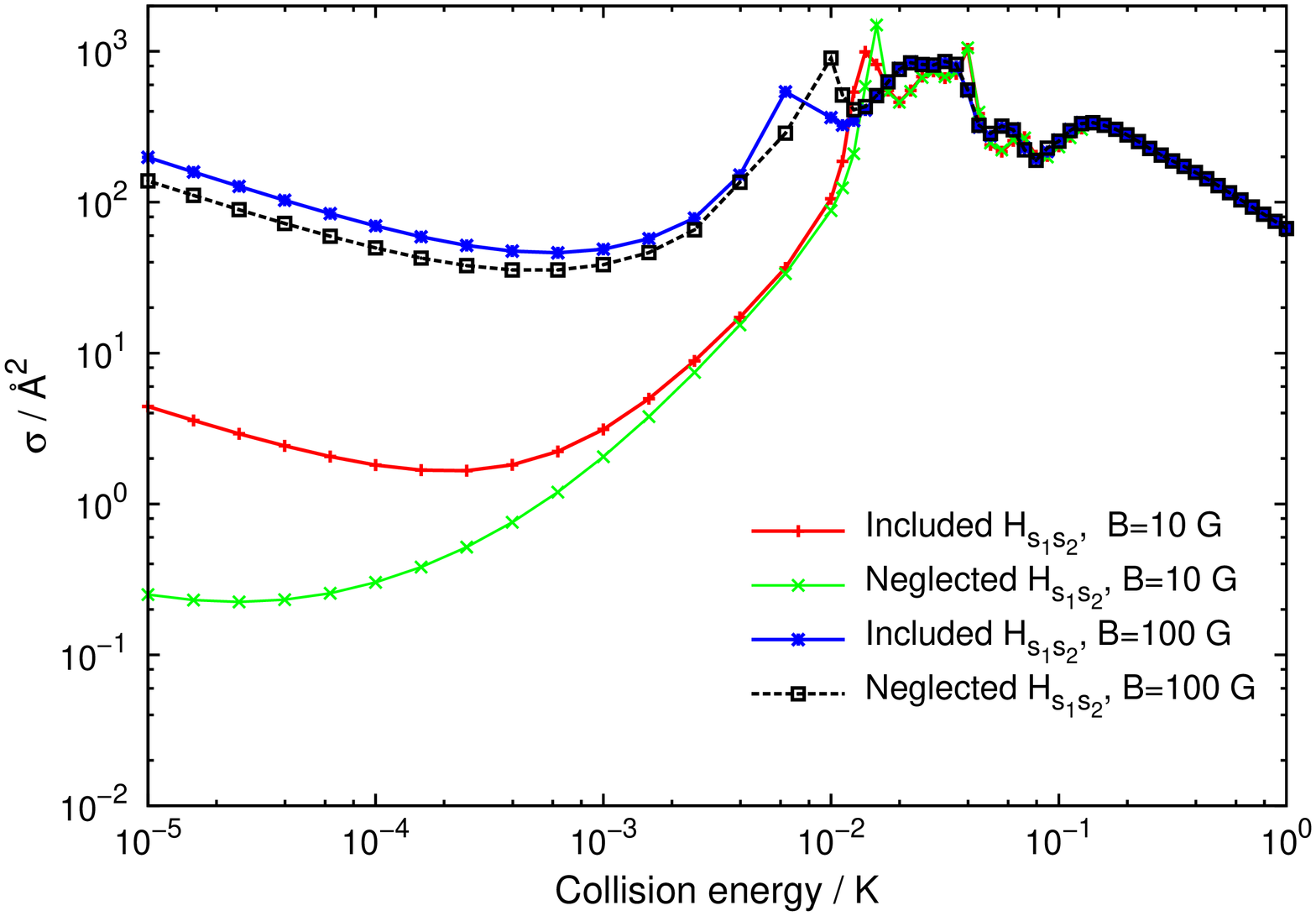}}\\ \hspace*{-0.20cm}
{\includegraphics[scale=0.32,angle=-90]{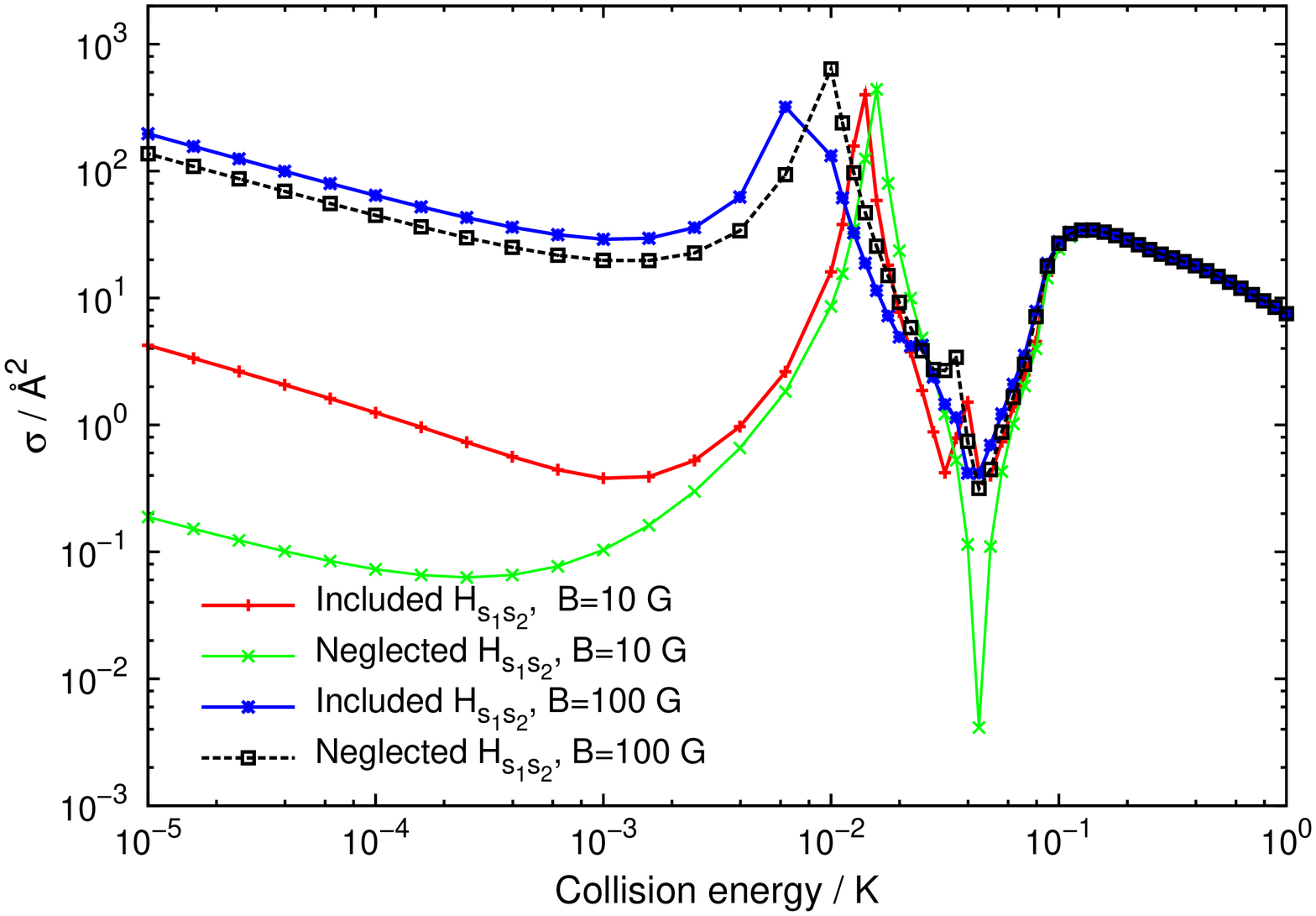}}
\caption{
Comparison of the total inelastic cross sections (upper panel) and the
$s$-wave contributions to them (lower panel) for N+OH, obtained with
the spin-spin dipolar interaction included or neglected in the
Hamiltonian, for magnetic fields $B=10$ and 100 G. } \label{fig4}
\end{figure}
Fig.\ \ref{fig3} shows the cross sections for Zeeman relaxation in
collisions of OH($X^2\Pi$$,|m_j=3/2,e\rangle$) with
N($^4$S,$\;|m_{s_1}=3/2\rangle$) for magnetic field strengths $B=10$,
100, 500 and 1000 G. At low collision energies (below 0.1 mK), the
cross sections behave according to the Wigner threshold laws
\cite{wigner}: the elastic cross section is constant, while the total
inelastic cross section grows with decreasing energy as $E^{-1/2}$. The
elastic cross section is almost unaffected by the magnetic field. At
ultralow collision energies, the inelastic cross sections are
suppressed due to centrifugal barriers in the outgoing channels, and
the total inelastic cross section grows with increasing field because
the increasing kinetic energy release helps overcome these barriers.
For example, the energy released by relaxation to the state
$|m_{s_1}=3/2\rangle|m_j=-1/2,e\rangle$ at a field of 560 G is
sufficient to overcome the $d$-wave barrier, and Fig.\ \ref{fig3} shows
how the inelastic cross section is enhanced for fields of 500 G and
higher in the $s$-wave regime. For collision energies between 4 mK and
80 mK, both the elastic and inelastic cross sections feature numerous
resonances, mostly Feshbach resonances due to coupling with
higher-energy closed channels.

\begin{figure*}[t!]
\centering
\begin{tabular}{cc}
\vspace*{0cm}
{\hspace*{0cm}\includegraphics[scale=0.33,angle=-90]{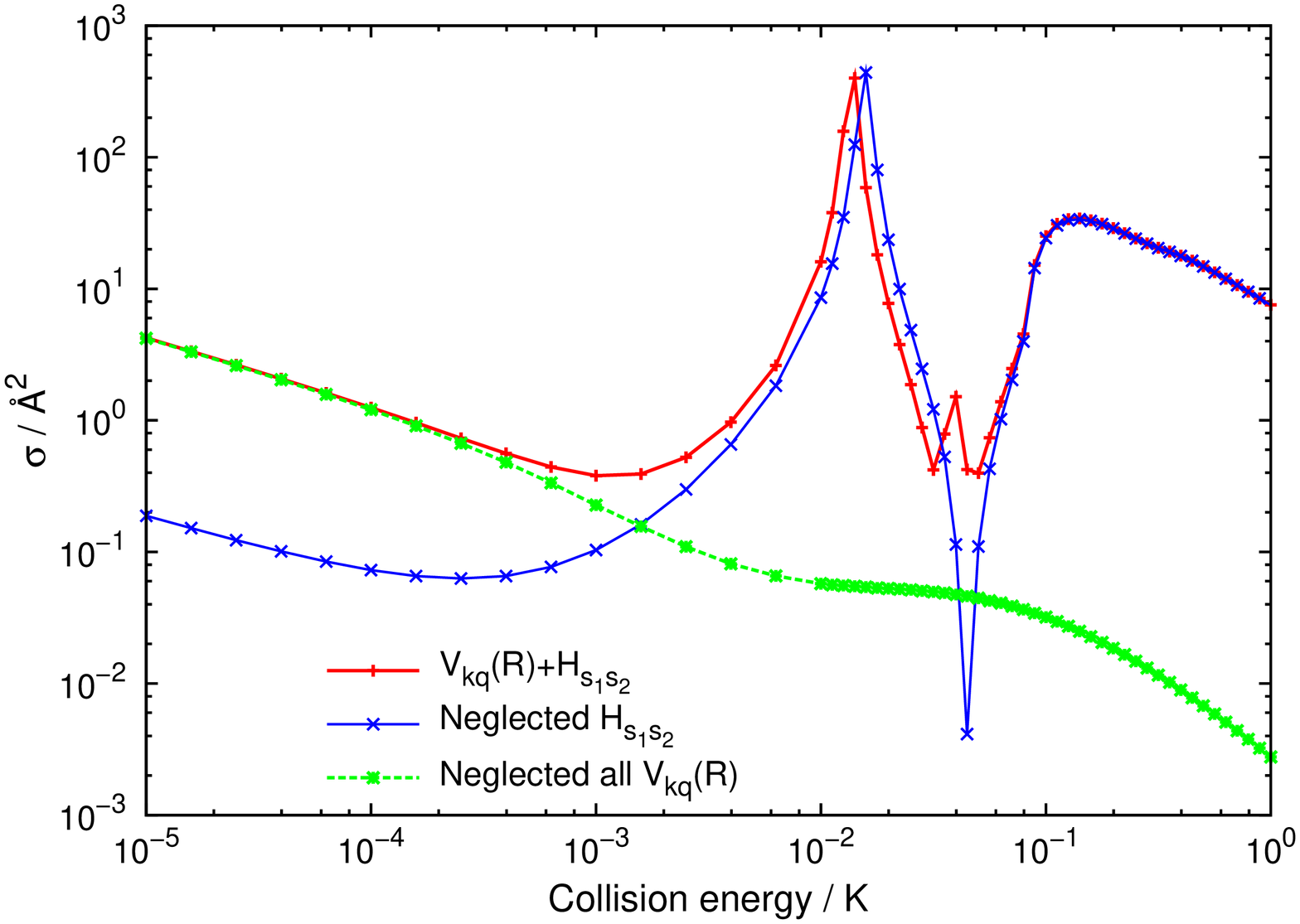}}&
\includegraphics[scale=0.33,angle=-90]{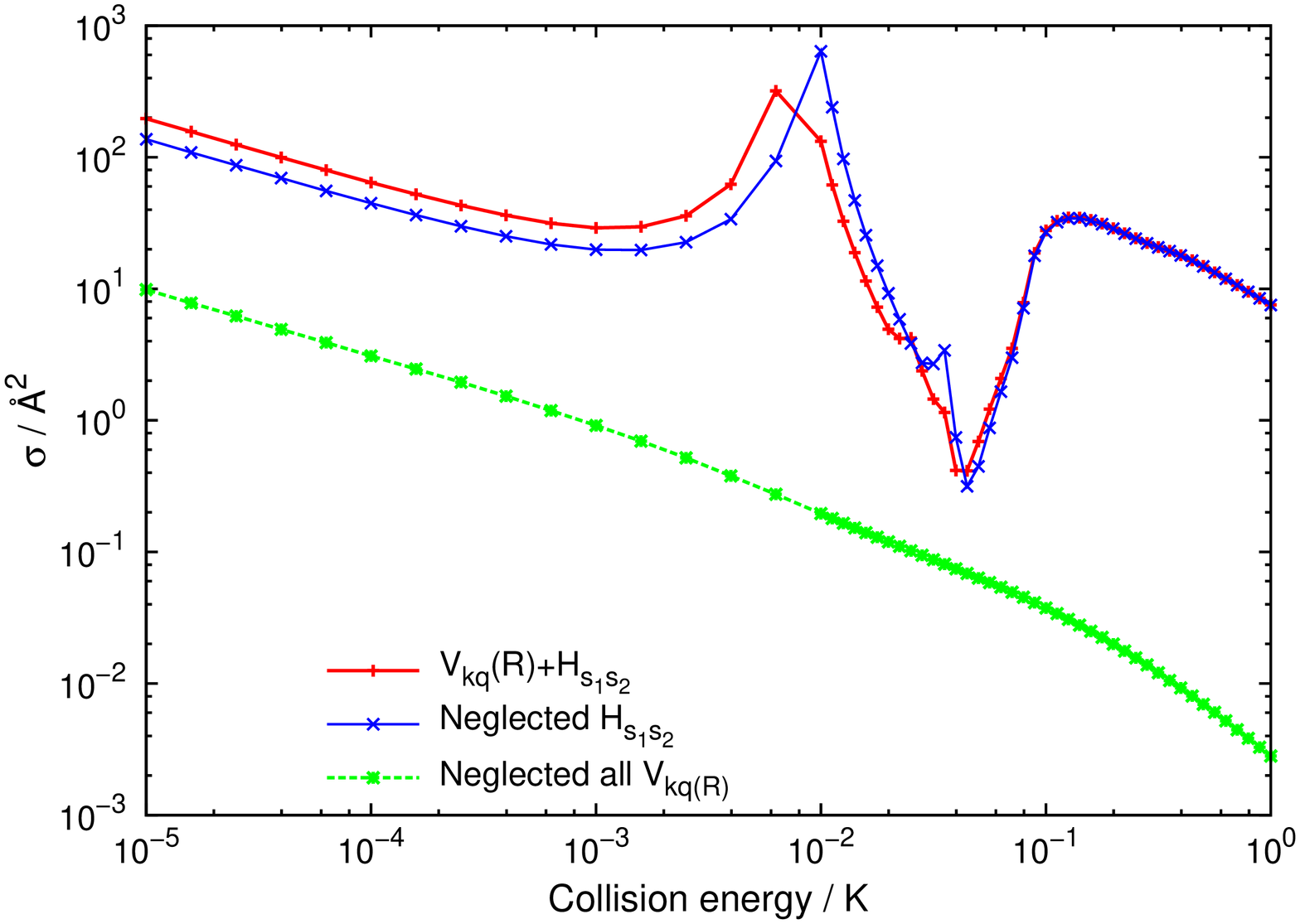}\\
\end{tabular}
\caption{
Comparison of the $s$ wave total inelastic cross sections for N+OH with
those obtained with either the spin-spin dipolar term or the anisotropy
of the interaction potential neglected. Left-hand panel: $B=10$ G; right-hand
panel: $B=100$ G. } \label{fig6}
\end{figure*}

\begin{figure}[b!]
\centering
\vspace*{0cm}
{\hspace*{0cm}\includegraphics[scale=0.70,angle=0]{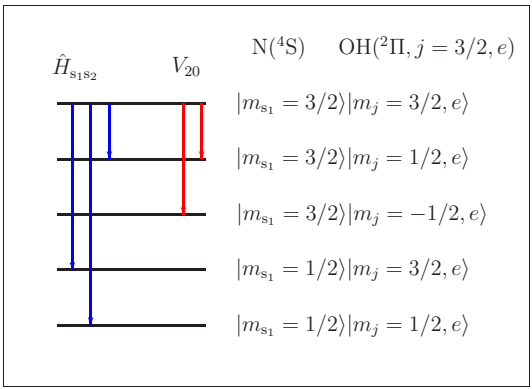}}\\
\caption{
Pattern of first-order couplings between different Zeeman levels of
N($^4$S)+OH($X^2\Pi,j=3/2$) through the spin-spin dipolar interaction
and the anisotropy of the interaction potential, for incoming $s$ wave
($L_i=0$) and outgoing $d$ wave ($L_f=2$). } \label{fig5}
\end{figure}

As discussed above, there are two mechanisms driving transitions
between different Zeeman levels, one driven by the spin-spin dipolar
term $\hat{H}_\mathrm{s_1s_2}$ and the other driven by the anisotropy
of the intermolecular potential $V_{kq}(R)$. The mechanism involving
$\hat{H}_\mathrm{s_1s_2}$ dominates for low fields (10 G and below) and
in the $s$-wave regime. For higher fields (100 G and above), the
opposite is true and the relaxation is driven by $V_{kq}(R)$. Fig.\
\ref{fig4} shows the integral cross sections and the $s$-wave
contribution for the two lowest fields (10 and 100 G), with the
$\hat{H}_\mathrm{s_1s_2}$ term in the Hamiltonian included or
neglected. Fig.\ \ref{fig6} compares the $s$-wave contributions for the
same two fields with those obtained by neglecting either the spin-spin
dipolar term $\hat{H}_\mathrm{s_1s_2}$ or all the anisotropic terms
$V_{kq}(R)$. At 10 G, $\hat{H}_\mathrm{s_1s_2}$ greatly enhances
inelastic processes in the ultracold regime: at $10^{-5}$ K, the
enhancement is almost two orders of magnitude. However, for $B=100$ G,
$V_{kq}(R)$ is dominant over the whole range of energies. 


The way in which the spin-spin dipolar interaction induces transitions
between different Zeeman levels is exactly parallel to that described
by Janssen {\em et al.} \cite{liesbeth2,liesbeth3}. It is a purely long-range
effect caused by narrowly avoided crossings between the potential
adiabats at very long range, which enable transitions between Zeeman
levels without the need to penetrate centrifugal barriers. In the
present case, avoided crossings due to the dipolar term are present
between the adiabat asymptotically correlating with the incident
$s$-wave channel $|m_{s_1}=3/2\rangle|m_j=3/2,e\rangle$ and other
adiabats correlating with the states
$|m_{s_1}=3/2\rangle|m_j=1/2,e\rangle$,
$|m_{s_1}=1/2\rangle|m_j=3/2,e\rangle$, and
$|m_{s_1}=1/2\rangle|m_j=1/2,e\rangle$. The $p$-wave and higher-$L$
contributions to the total inelastic cross sections are almost
unaffected by the inclusion of $\hat{H}_\mathrm{s_1s_2}$ for any field
and collision energy. This arises because the long-range avoided
crossings for incident channels with centrifugal barriers are
energetically inaccessible at low energies.

Channels corresponding to different Zeeman levels are also directly
coupled by the anisotropy of the intermolecular potential $V_{kq}(R)$.
Fig.\ \ref{fig5} shows a schematic illustration of the first-order
couplings by $\hat{H}_\mathrm{s_1s_2}$ and $V_{kq}(R)$ for collisions
involving an incoming $s$ wave and outgoing $d$ waves. Because of this,
long-range avoided crossings are present even if we neglect
$\hat{H}_\mathrm{s_1s_2}$. However, the effect of the avoided crossings
on collision outcomes is much more pronounced  for crossings due to
$\hat{H}_\mathrm{s_1s_2}$ than for those due to $V_{kq}(R)$. The latter
dies off much faster with $R$ than $\hat{H}_\mathrm{s_1s_2}$, and is
one or two orders of magnitude weaker at the positions of the
long-range avoided crossings. The ratio of the coupling strengths is
approximately $\hat{H}_\mathrm{s_1s_2}(R)/V_{20}(R)=E_{\rm
h}\alpha^2/C_{6}^{20} (R/a_0)^{-3}$. The avoided crossings for $B=10$ G
occur at distances ranging from 159 to 342 $a_0$, corresponding to a
ratio $\hat{H}_\mathrm{s_1s_2}(R)/V_{20}(R)$ between 10 and 100. It
follows from an approximate Landau-Zener model \cite{drake} that the
probability of ending in a different asymptotic level after a
nonadiabatic transition is proportional to the square of the coupling
between the diabats if the coupling is relatively small.
\begin{figure}[t!]
\vspace*{0cm}
{\hspace*{0cm}\includegraphics[scale=0.33,angle=-90]{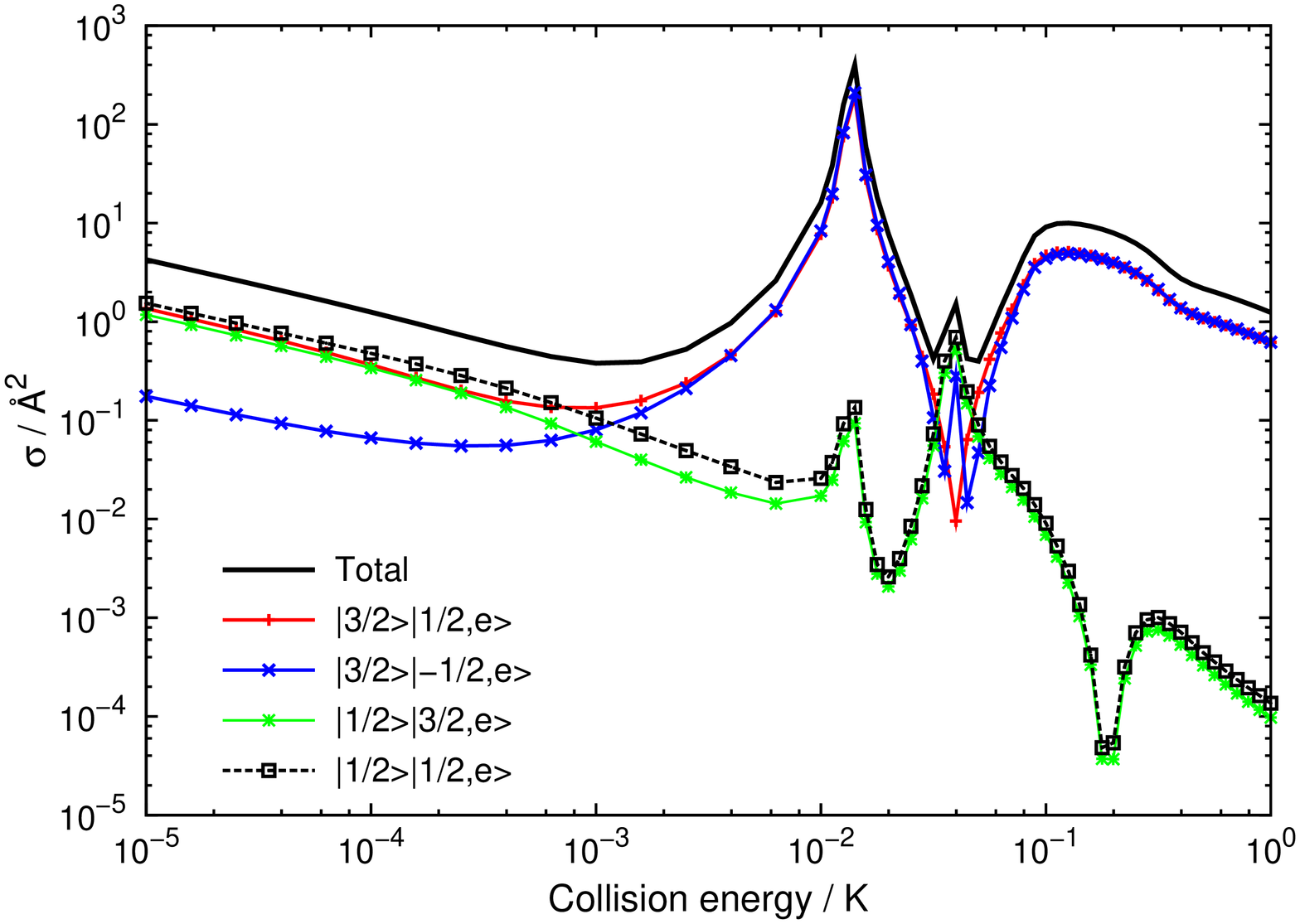}}\\
\includegraphics[scale=0.33,angle=-90]{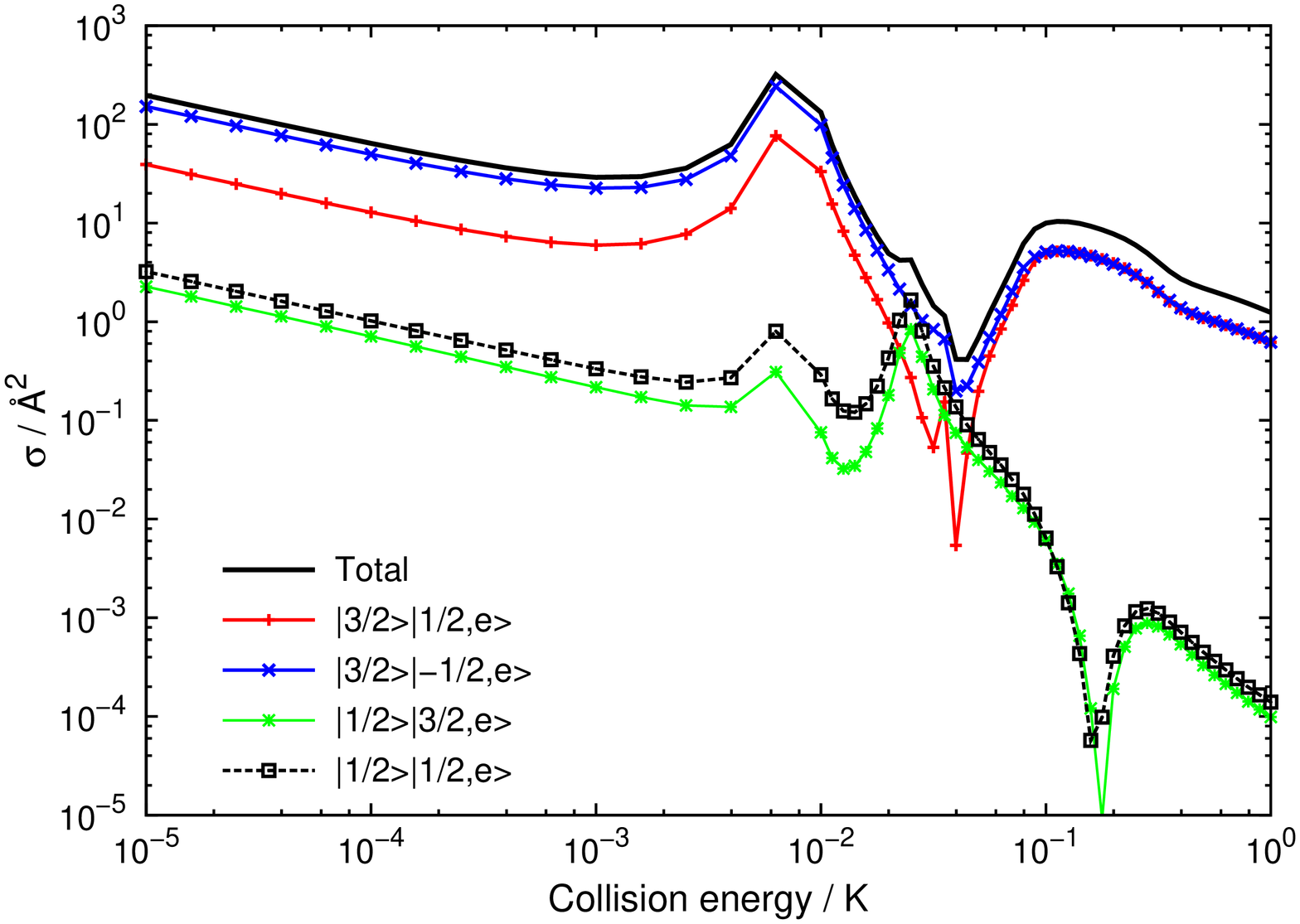}
\caption{
State-to-state inelastic cross sections ($s$-wave contribution only) for
fields of 10 G (upper panel) and 100 G (lower panel).
}
\label{fig7}
\end{figure}
\begin{figure}[t!]
\vspace*{0cm}
{\hspace*{0.4cm}\includegraphics[scale=0.35,angle=-90]{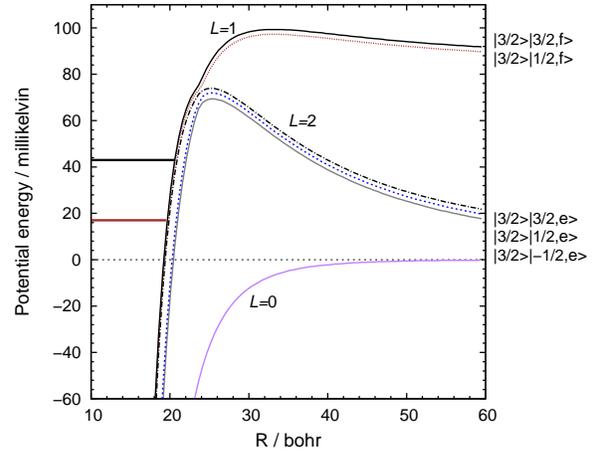}}\\
\caption{
The lowest adiabatic potential energy curves for $M_{\rm tot}=3$ and
$L_{\rm max}=2$ correlating with thresholds with the state of the N
atom unchanged ($|m_{s_1}=3/2\rangle$), at $B=10$ G. Two solid
horizontal lines indicates the position of the bound states responsible
for the two sharp Feshbach resonances in the $s$ and $d$-wave
contribution to the inelastic cross sections.} \label{fig10}
\end{figure}

The interplay between the spin-spin dipolar term and the intermolecular
potential terms is also manifested in the state-to-state cross
sections. Fig.\ \ref{fig7} shows state-to-state cross sections
($s$-wave contributions only) for $B=10$ G and 100 G. At $B=10$ G, in
the region where the $\hat{H}_\mathrm{s_1s_2}$ term dominates (below 1
mK), the most important transitions are to states with $m_j$ or
$m_{s_1}$ quantum numbers reduced by 1, which are those coupled to the
incident channel $|m_{s_1}=3/2\rangle|m_j=3/2,e\rangle$ by
$\hat{H}_\mathrm{s_1s_2}$, while for collision energies above 1 mK the
dominant inelastic channels become
$|m_{s_1}=3/2\rangle|m_j=1/2,e\rangle$ and
$|m_{s_1}=3/2\rangle|m_j=-1/2,e\rangle$, which are those coupled by
$V_{kq}(R)$. At $B=100$ G, only channels coupled by $V_{kq}(R)$ are
important.

The $s$-wave cross sections at $B=10$ G exhibit two distinct resonant
structures: a strong feature near 15 mK and a weaker one around 41 mK.
Both are Feshbach resonances caused by coupling to closed channels
arising from the $f$ component of the $\Lambda$ doublet of OH. The
coupling arises almost exclusively from the $V_{10}(R)$ term in the
intermolecular potential, which couples states of different monomer
parity. The Feshbach resonance near 15 mK can be attributed to a bound
state on the $p$-wave adiabat correlating with the
$|m_{s_1}=3/2\rangle|m_j=1/2,f\rangle$ threshold, as shown in Fig.\
\ref{fig10}. This resonance moves to smaller energies with increasing
field, because the energy difference between the $|m_j=3/2,e\rangle$
and $|m_j=1/2,f\rangle$ states (red and dotted blue lines in Fig.\
\ref{fig2}, respectively) decreases as the field increases. For
sufficiently large field ($B>1200$ G), this resonance will disappear as
the $|m_j=1/2,f\rangle$ level drops below $|m_j=3/2,e\rangle$. The
second Feshbach resonance near 41 mK can be attributed to a bound state
on the $p$-wave adiabat correlating with the
$|m_{s_1}=3/2\rangle|m_j=3/2,f\rangle$ threshold. The position of this
resonance is almost unaffected by the field strength since the energy
difference between the two spin-stretched states, $|m_j=3/2,e\rangle$
and $|m_j=3/2,f\rangle$, is independent of magnetic field.
\begin{figure}[t!]
\vspace*{0cm}
{\hspace*{0cm}\includegraphics[scale=0.35,angle=-90]{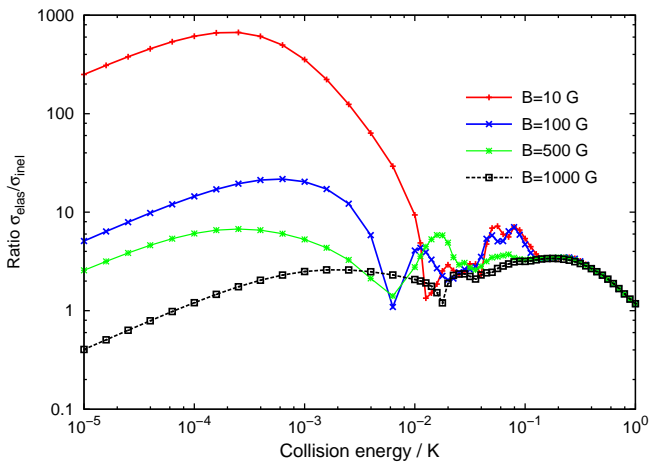}}\\
\caption{
Ratio of elastic to total inelastic cross sections for N+OH at different magnetic fields.
}
\label{fig8}
\end{figure}

The fact that these are Feshbach (rather than shape) resonances is
confirmed by several observations. First, the $d$-wave contributions to
the inelastic cross sections show resonant structures at exactly the
same energies as the $s$-wave contribution. Secondly, the positions and
shapes of the Feshbach resonances can be reproduced using even the
smallest possible basis set that allows inelastic transitions, with
$j\le3/2$, $L\le2$, and potential terms $V_{kq}(R)$, $k\le2$. Thirdly,
the presence of the $V_{10}(R)$ term, which does not couple the
incident and outgoing channels directly, is crucial for the existence
of the resonances. It is worth noting that no such structure due to
Feshbach resonances would be present for collisions involving the
initial state $|m_{s_1}=3/2\rangle|m_j=3/2,f\rangle$, with OH in the
upper component of its $\Lambda$ doublet, since no low-lying closed
channels are present in that case. However, molecules in the $f$ state
are likely to undergo fast relaxation to the $e$ state in collisions
driven directly by $V_{10}(R)$.

Fig.\ \ref{fig8} shows the ratio of the elastic to total inelastic
cross sections as a function of collision energy. The ratio is not
favourable for sympathetic cooling of OH by collision with ultracold N
atoms, except at fields below 10 G and collision energies below 1 mK.
The cross sections presented here may be compared to those for
N($^4$S)+NH($^3\Sigma^-$) by \.Zuchowski and Hutson \cite{pzuch:n}. The
ratio of the elastic to inelastic cross sections is at least an order
of magnitude lower for N+OH than for N+NH. Two main reasons for this
can be identified. First, the spin-stretched component of the
rotational ground state of NH ($^3\Sigma^-$, $n=0$) is not directly
coupled by the potential anisotropy to any other accessible Zeeman
level, whereas such a coupling does exist for the ground state of
OH($^2\Pi,j=3/2$) (or any other molecule with $j\ge1$). Secondly, there
are low-lying states arising from the $f$ component of the $\Lambda$
doublet in the OH radical that create many Feshbach resonances and
increase the inelasticity. Both effects are particularly strong for
collision energies above 10 mK, where the contributions from $p$ and
$d$ incoming waves to the inelastic cross sections are dominant. For
all field strengths, the ratio of elastic to inelastic cross sections
at collision energies above 1 mK is more than 10 times larger for N+NH
than for N+OH.

\subsection{Potential dependence}
\begin{figure}[b!]
\vspace*{-0cm}
{\hspace*{0cm}\includegraphics[scale=0.33,angle=-90]{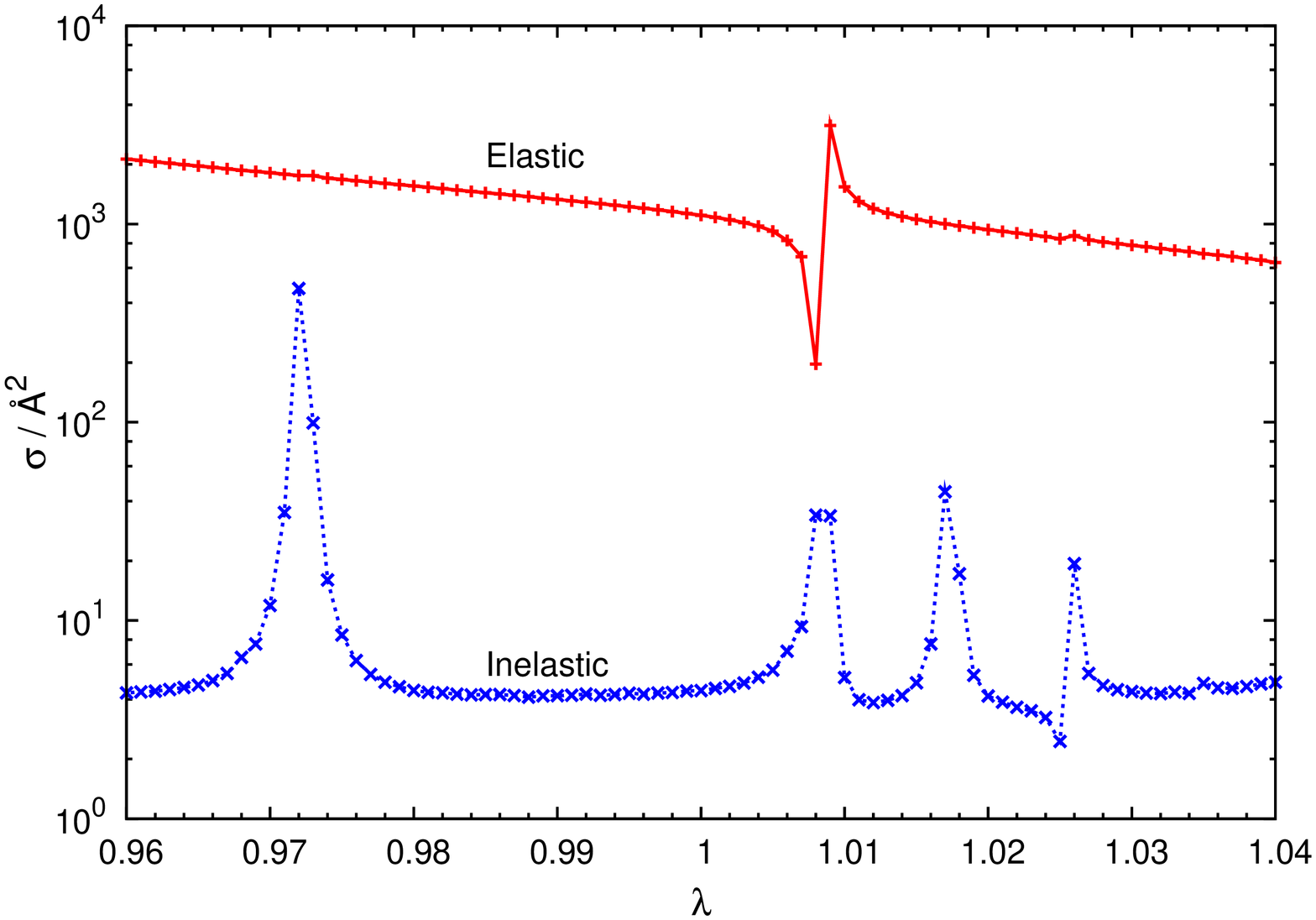}}\\
\includegraphics[scale=0.33,angle=-90]{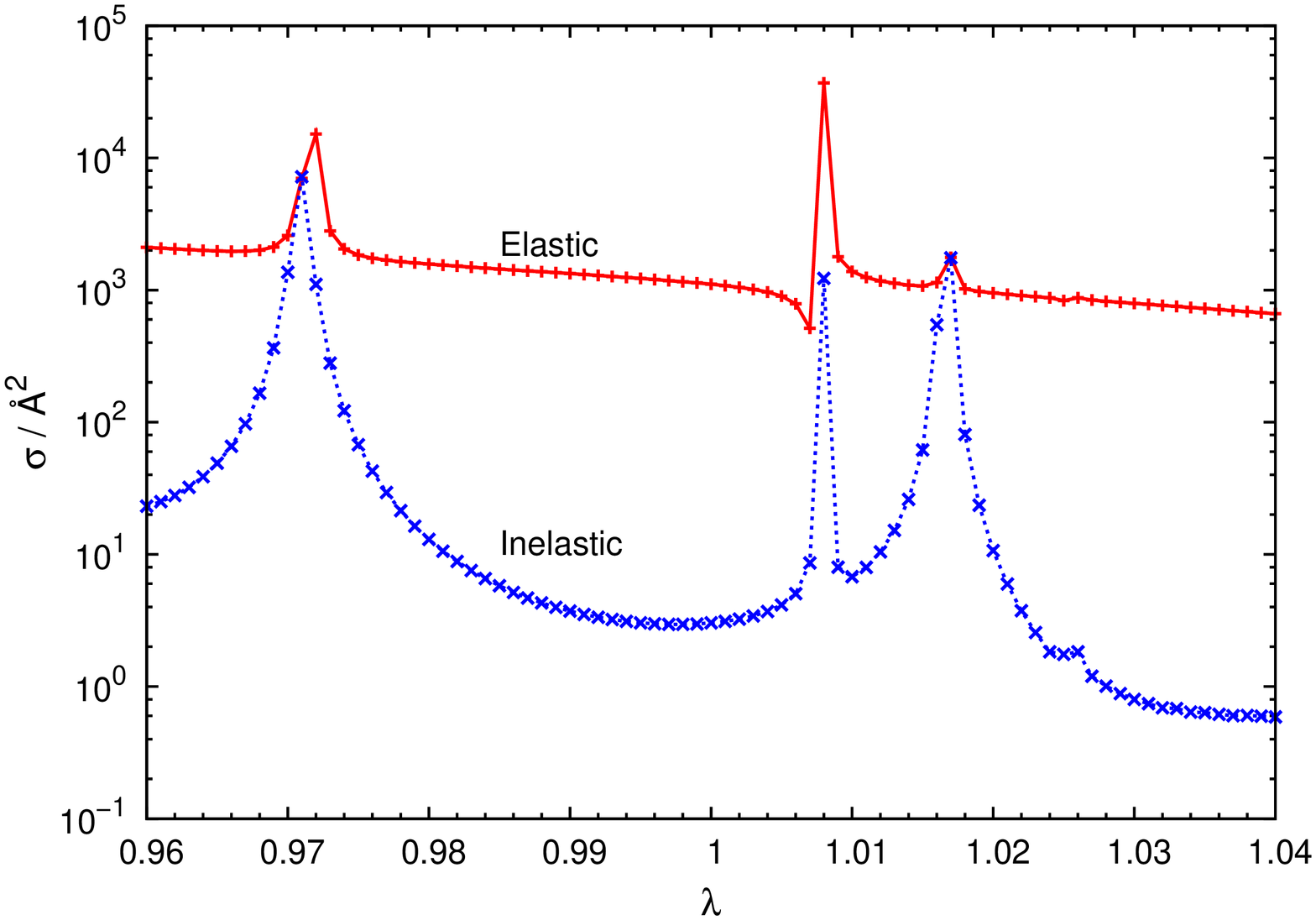}
\caption{
Cross sections obtained with the interaction potential
scaled by a constant factor, $V(R)\to \lambda\cdot V(R)$, for
magnetic field $B=10$ G and collision energies of 10 $\mu$K (upper panel)
and 1 mK (lower panel).
}
\label{fig9}
\end{figure}

The results of scattering calculations at ultralow collision energies
are in general very sensitive to the details of interaction potentials.
To estimate the accuracy of the calculated potential energy surfaces
for N+OH, we have carried out additional electronic structure
calculations for the geometry corresponding to the global minimum of
the potentials at the linear N--OH geometry. In the aug-cc-pV5Z basis
set, the global minimum has a well depth of 120.9 cm$^{-1}$, while in
the aug-cc-pV6Z basis set this shifts to 121.8 cm$^{-1}$. Based on
these two results, we can estimate the complete basis set limit of the
CCSD(T) method to be 122.9 cm$^{-1}$, using the extrapolation formula
for correlation energy as given in Ref. \cite{Bak:2001}. This corresponds to an error
estimate of 1.7\% for our full potential surfaces using the aug-cc-pV5Z
basis set. To estimate the error in the correlation energy obtained
from the CCSD(T) method, we have performed full
configuration-interaction (FCI) calculations with eight electrons
correlated in the cc-pVDZ basis set. The relative contribution of the
FCI correction to the CCSD(T) result should only be weakly dependent on
the basis set used, so even in the small cc-pVDZ basis set we should
obtain a reliable estimate of the FCI valence-valence correlation
correction. The FCI correction to the CCSD(T) result accounts for
approximately 1.5\% of the interaction energy at the global minimum. We
can thus estimate the uncertainty of our potential energy surfaces to
be 4\% at worst.

To assess the sensitivity of the scattering results to the uncertainty
in the interaction potential, we have carried out calculations with the
interaction potential scaled by a constant factor $\lambda$ in the
range $0.96\leq\lambda\leq1.04$, corresponding to the estimated error
bounds in the calculated potential energy surfaces. The results at
$B=10$ G are shown in Fig.\ \ref{fig9} for collision energies of 1 mK
and 10 $\mu$K. The weak dependence of the cross sections on the
potential scaling is disturbed by the presence of sharp resonances,
which occur when bound states of the N-OH complex cross the incoming
threshold (or more precisely the collision energy) as a function of
$\lambda$. Two of the peaks in the inelastic cross sections, near
$\lambda=1.010$ and $\lambda=1.026$, can be attributed to the Feshbach
resonances in the $s$ and $d$ partial-wave contributions discussed
above. The two additional peaks at $\lambda=0.97$ and $\lambda=1.017$,
which broaden substantially with collision energy, are due to shape
resonances in the $p$-wave partial cross section. If the true potential
happens to bring one of these resonances close to zero energy, it may
change the ratio of elastic to inelastic cross sections quite
dramatically. However, Fig.\ \ref{fig9} shows that the resonances occur
in quite narrow ranges of $\lambda$, so that there is a low probability
that the true potential will be such that the ratio of elastic to
inelastic cross sections is seriously affected by resonances for
collision energies below 1 mK. It may also be noted that the numerical
results obtained with the unscaled potential ($\lambda=1$) are fairly
typical of the range expected for N+OH on plausible interaction
potentials, in the sense that the low-energy elastic cross section
(around 1000 \AA$^2$) is close to the value $\sigma=4\pi\bar{a}^2=712$
\AA$^2$ obtained from the mean scattering length $\bar{a}$ defined by
Gribakin and Flambaum \cite{Gribakin:1993}.

\section{Summary and conclusions}

We have presented a theoretical study of the relaxation processes in
collisions between an atom in an open-shell S state and a molecule in a
$^2\Pi$ state, in a magnetic field, using the example of
N($^4$S)+OH($^2\Pi$). The transitions between different Zeeman levels
in such collisions are driven by two mechanisms: coupling through the
spin-spin dipolar term and through the anisotropy of the interaction
potential. Both mechanisms are present in first order. The spin-spin
dipolar term dominates when both the collision energy and the magnetic
field are low, while the anisotropy of the interaction potential
dominates at higher energies or fields. In the latter regime, the
spin-spin dipolar term can be neglected. Neglecting the dipolar
interaction is equivalent to treating the atom as closed-shell, which
dramatically reduces the cost of the scattering calculations.

An important general point is that spin relaxation collisions can be
driven directly by the anisotropy of the interaction potential for any
molecule that has rotational angular momentum. Since the anisotropies
of atom-molecule and molecule-molecule interaction potentials are
typically quite large, this will often provide an important trap loss
mechanism for such states. For molecules in $^2\Pi$ states, this is
true even for the molecular ground state.

For the case of N+OH, the spin-spin dipolar term dominates at collision
energies below about 1 mK and magnetic fields of 10 G or less. In this
regime, the ratio of elastic to inelastic cross sections is greater
than 100 and thus favourable for sympathetic cooling. However, if
either the collision energy or the magnetic field is significantly
above this, inelastic processes due to the potential anisotropy
dominate and the ratio of elastic to inelastic cross sections falls.
This suggests that sympathetic cooling of OH by collisions with N atoms
is unlikely to be successful except at collision energies below 1 mK.

\section*{Acknowledgments}
The authors are grateful to the Polish Ministry of Science and Higher
Education (project N N204 215539) and to the UK Engineering and
Physical Sciences Research Council for financial support. The
collaboration was supported by the EuroQUAM Programme of the European
Science Foundation.

\footnotesize{
\bibliography{noh_cold,all} 
\bibliographystyle{rsc} 
}

\end{document}